\documentclass[12pt]{article}%
\usepackage{bm} %Letras griegas en negrita
\usepackage{amssymb,amsmath,amsthm}
\usepackage{graphicx}
\usepackage[height=9in,left=1in,right=0.75in,bottom=1in]{geometry}
\usepackage[breaklinks=true,bookmarksopen=true,colorlinks=false]{hyperref}
\usepackage[super]{natbib}
\usepackage{xcolor}
\usepackage{amsmath}%
\usepackage[title]{appendix} %titulo en los apendices
\usepackage{comment} %Comentar bloques
\usepackage{thmtools} %Para mejor manejo de los teoremas y poder continuarlos
\usepackage{bold-extra} %Small caps y bold al mismo tiempo
\usepackage{enumitem} %Enumeraciones con labels, etc.
\usepackage{graphicx} %Figures in a row
\usepackage{rotating} %Pagina apaisada
\usepackage{mathtools}
\usepackage[font = footnotesize]{caption} %Modificar los captions de las figuras
\usepackage{booktabs}
\usepackage{multirow}
\usepackage{subcaption}

\RequirePackage{hypernat} %Hyperref y bibliografia

\DeclareCaptionLabelFormat{bsc}{\textbf{\textsc{#1}\ #2}}
\newcommand*{\E}{\operatorname{E}}
\newcommand*{\prob}{\operatorname{P}}
\newcommand*{\ind}{\mathbf{1}}

\declaretheoremstyle[
spaceabove=10pt, spacebelow=10pt,
headfont=\normalfont\scshape\bfseries,
notefont=\normalfont\scshape,
notebraces={(}{)},
bodyfont=\itshape,
headpunct=,
postheadspace=1em
]{THMS}

\declaretheoremstyle[
spaceabove=6pt, spacebelow=6pt,
headfont=\normalfont\scshape\bfseries,
notefont=\normalfont\scshape,
notebraces={(}{)},
bodyfont=\normalfont,
headpunct=,
postheadspace=1em
]{DEF}

\declaretheoremstyle[
spaceabove=12pt, spacebelow=6pt,
headfont=\normalfont\scshape\bfseries,
notefont=\normalfont\scshape,
notebraces={(}{)},
bodyfont=\normalfont,
headpunct=,
postheadspace=1em,
qed=$\blacksquare$
]{EX}

\declaretheoremstyle[
spaceabove=12pt, spacebelow=12pt,
headfont=\normalfont\scshape\bfseries,
notefont=\normalfont\scshape\bfseries,
notebraces={}{},
bodyfont=\normalfont,
headpunct={:},
qed=$\blacksquare$,
postheadspace=1em
]{PROOF}
\declaretheorem[style=THMS, numberwithin=section, name=Theorem]{thm}
\declaretheorem[style=THMS, numberwithin=section, name=Proposition]{prop}
\declaretheorem[style=THMS, numberwithin=section, name=Lemma]{lma}

\declaretheorem[style=PROOF, numbered=no, name=Proof of ]{proofc}
\declaretheorem[style=PROOF, numbered=no, name=Proof]{proofd}

\numberwithin{equation}{section}
\mathtoolsset{showonlyrefs}

\begin{document}

\title{Dynamic prediction of death risk given a renewal hospitalization process}
\author{
Telmo J. P\'{e}rez-Izquierdo\\ \textit{Department of Economic Analysis, University of the Basque Country UPV/EHU} \\\textit{BCAM - Basque Center for Applied Mathematics}
\and
Irantzu Barrio \\ \textit{Department of Mathematics, University of the Basque Country UPV/EHU} \\\textit{BCAM - Basque Center for Applied Mathematics}
\and
Cristóbal Esteban \\ \textit{Pneumology Service, University Hospital of Galdakao}
}
\date{\today}
\maketitle

\begin{abstract}
	Predicting the risk of death for chronic patients is highly valuable for informed medical decision-making. This paper proposes a general framework for dynamic prediction of the risk of death of a patient given her hospitalization history, which is generally available to physicians. Predictions are based on a joint model for the death and hospitalization processes, thereby avoiding the potential bias arising from selection of survivors. The framework accommodates various submodels for the hospitalization process. In particular, we study prediction of the risk of death in a renewal model for hospitalizations ---a common approach to recurrent event modelling. In the renewal model, the distribution of hospitalizations throughout the follow-up period impacts the risk of death. This result differs from the prediction of death in the Poisson model for hospitalization, previously studied, where only the number of hospitalizations matters. We apply our methodology to a prospective, observational cohort study of 512 patients treated for COPD in one of six outpatient respiratory clinics run by the Respiratory Service of Galdakao University Hospital, with a median follow-up of 4.7 years. We find that more concentrated hospitalizations increase the risk of death.

	\vspace{5pt}

\end{abstract}

%%%%%%%%%%%%%%%%%%%%%%%%%%%%%%%%%%%%%%%%%%%%%%%%%%%%%%%%%%%%%%%%%%%%%%%%%%
%%%%%%%%%%%%%%%%%%%%%%%%%%%%%%%%%%%%%%%%%%%%%%%%%%%%%%%%%%%%%%%%%%%%%%%%%%
\newpage
\section{Introduction}
\label{sec:intro}

In numerous clinical studies, participants are regularly monitored, and response data, such as biomarkers, vital signs, or hospitalizations, are collected. Additionally, there is often a focus on the time until a specific event occurs, such as death. The longitudinal data collected may be censored by this time-to-event outcome.  Therefore, modelling the longitudinal and event-time outcomes separately can lead to biased effect size estimates if the two processes are correlated.\citep{ibrahim2010basic} Joint modelling of longitudinal and time-to-event data has been a hot topic of research in recent years.\citep{rizopoulos2012joint, asar2015joint, hickey2016joint} Furthermore, joint models have been proposed for modelling a recurrent event and a terminal event.\citep{liu2004shared,rondeau2007joint} %\textcolor{red}{Citas a las aplicaciones?}

%For instance, they have been used to model the serum digoxin levels and heart-failure mortality \citep{field2023joint}; to study the physical activity patterns by jointly modeling the intensity (activity count) and timing (timestamp) from accelerometers measurements \citep{geraci2023joint}, or to evaluate the treatment effect in a  in chronic obstructive pulmonary disease (COPD) clinical trials by quantifying the association between longitudinal improvements in forced expiratory volume in one second (FEV1) and exacerbation risk reduction \citep{zhudenkov2021longitudinal}.

Joint models mitigate the bias in the estimates of covariate effects and allow to quantify and clinically interpret the dependence between both processes. Moreover, joint models provide an outstanding prediction framework, as the paths of both processes ---on top of their dependence--- is modelled. This paper provides a general dynamic prediction result for the risk of the terminal event given the history of recurrent events. By dynamic we mean that the prediction can be updated throughout the patient's follow-up, as more data becomes available. The result is general in the sense that it allows for arbitrary recurrent event submodels.

Our development is motivated by the study of the relationship between hospitalizations (the recurrent event) and death for Chronic Obstructive Pulmonary Disease (COPD) patients. \citet{suissa2012long} studied this relationship using a Cox model for the hazard of death given the number of hospitalizations. Nevertheless, the relationship between death and hospitalization processes has also been studied using joint models in other contexts.\citep{liu2004shared,gonzalez2005sex} We believe that this problem is particularly suitable for dynamic prediction as the history of hospitalization is often recorded for chronic patients. Thus, incorporating this information for death risk prediction may enhance medical decision-making.

We extend the prediction results in \citet{mauguen2013dynamic}, which are valid for a Poisson model for hospitalizations (also known as calendar time model), to a general hospitalization submodel. This general model encompasses, among others, the Poisson and the renewal model (also known as gap time model) for the hospitalization process (or any type of recurrent events). The renewal model is particularly relevant, as it has usually been considered for the hospitalization process.\citep{gonzalez2005sex,louzada2017gap,suissa2012long} We thoroughly study the dependence of the risk of death given the history of hospitalizations for the renewal model. We find that the distribution of hospitalizations during the patient's follow-up ---that is, the timings of the hospitalizations--- determines the risk of death. This contrasts with the Poisson model, where only the number of hospitalizations during follow-up matters when predicting the risk of death.\citep{mauguen2013dynamic} Hence, the renewal model then resembles the clinicians' decision-making process, where both the number of hospitalizations and the gap-time between hospitalizations are clinically important factors.

This paper is organized as follows. Section~\ref{sec:data} introduces the motivating case: the study of risk of death of COPD patients. Section~\ref{sec:methods} presents the methodological contributions of the paper. In Section~\ref{sec:analysis} we apply our results to predict the risk of death, given the hospitalization history, for COPD patients. Section~\ref{sec:discussion} concludes with a discussion. A more detailed technical discussion of the methodological contributions and the results is presented in the Appendices.

%%%%%%%%%%%%%%%%%%%%%%%%%%%%%%%%%%%%%%%%%%%%%%%%%%%%%%%%%%%%%%%%%%%%%%%%%%
%%%%%%%%%%%%%%%%%%%%%%%%%%%%%%%%%%%%%%%%%%%%%%%%%%%%%%%%%%%%%%%%%%%%%%%%%%
\section{Motivating study: Risk of death of COPD patients}
\label{sec:data}

%Benefits of joint models: The practical application of joint modeling in analyzing medical data, is increasing because (a) they provide more efficient estimates of the covariate effects on the time to event, (b) they provide more efficient estimates of the covariate effects of the longitudinal marker, (c) they reduce bias in the estimates of the overall covariate effect, and (d) they allow to quantify and clinically interpret the dependence between both processes. 

%Also some intuition about join modelling: conditional probs of death given hospitalization look weird. Selection of ``resilient'' individuals when conditioning on hospitalizations \citep[see Table~4 in][which shows that death hazard does not increase after the third hospitalization]{suissa2012long}.

COPD's influence on global health and healthcare policies is significant and expanding. It is currently one of the most paradigmatic chronic respiratory condition, with projections indicating further escalation in the years ahead.\citep{soriano2020prevalence} COPD accounted for 3\% of all deaths in 2021 across OECD countries,\citep{OECD_2023} and it was the third leading cause of death in 2019 according to the \href{https://www.who.int/news-room/fact-sheets/detail/chronic-obstructive-pulmonary-disease-(copd)}{World Health Organization March 2023 report}. The burden of disease will continue to increase, especially in women and regions with low to medium gross domestic product or income. \citep{boers2023global} The hospitalizations and readmission due to COPD exacerbation significantly impact healthcare utilization. COPD severe exacerbations  imply hospital admission, while patients exhibiting a pattern of hospitalizations lead to a deterioration of the patients' quality of life, which encompasses many dimensions of the patients' lives.\citep{esteban2020predictive} %\citep{lopez2022factors, guerrero2016readmission}. 

Several prediction models have been developed for COPD readmissions or mortality risk.\citep{quintana2022predictors, shah2022development, aramburu2019copd, arostegui2019computer}  Furthermore, the increasing hospitalization frequency has been shown to increase the risk of mortality.\citep{suissa2012long} Therefore, given the importance of these two outcomes (hospitalization and mortality due to COPD) and their intrinsic relationship, joint modeling of both processes is necessary for two reasons. On the one hand, it reduces the potential bias present in Cox regression models. The bias arises since the fact that long-lived patients tend to experience more hospitalizations could mask the accentuating effect of hospitalizations on the risk of death. On the other hand, it allows to evaluate the effect that hospitalization history (frequency and distribution) has on mortality and to make predictions of mortality risk based on history of hospitalizations during follow-up.

%the increasing exacerbation frequency has been shown to increase the risk of mortality \citep{suissa2012long}. This leads us to hypothesize that the increased frequency of hospital admissions has an impact on mortality in COPD patients. Therefore, given the importance of these two outcomes (hospitalization and mortality due to COPD) and their intrinsic relationship, joint modeling of both processes is necessary for two reasons. On the one hand, it allows us to model both processes together. On the other hand, we can evaluate the effect that hospitalization history (frequency and distribution) has on mortality.

In this work, we considered a prospective, observational cohort study of 512 patients recruited after being treated for COPD in one of six outpatient respiratory clinics run by the Respiratory Service of Galdakao University Hospital. Patients were consecutively included in the study if they had been diagnosed with COPD for at least 6 months and had been stable for at least 6 weeks. The protocol was approved by the Ethics and Research Committees of the hospital (reference 16/14). All candidate patients were given detailed information about the study, and all those included provided written informed consent. Sociodemographic, smoking habits, and clinical variables were recorded. Pulmonary function tests included forced spirometry and body plethysmography, and measurements of carbon monoxide diffusing capacity (DLCO). These tests were performed in accordance with the standards of the European Respiratory Society. 
\citep{ph1993lung} For theoretical values, we considered those of the European Community for Steel and Coal.\citep{stanojevic2017official} All variables were measured at baseline. The median follow-up time was 4.7 years (interquartile range: 2.66 -- 5.13 years). Patient hospitalizations were reviewed during the follow-up period.  A brief description of the main variables used for this study is presented in Table~\ref{tab:descriptive}. More detailed information regarding the dataset can be found elsewhere.\citep{esteban2024influence}

%Complete pulmonary function tests included forced expiratory volume in one second ($\text{FEV}_1$), body plethysmography, carbon monoxide diffusing capacity (DLCO) measurements, and respiratory muscle strength. These tests were performed in accordance with the standards of the Spanish Society of Respiratory Medicine and Thoracic Surgery (SEPAR) \citep{garcia2013spanish}. 

\begin{table}[h!]
	\centering
	\begin{tabular}{rcccc}
								   & Number & NA      & Median & Interquartile range \\ \hline
	Patients                       & 512     &     &        &                     \\
	Death events                   & 95 (18.5\%) &  &        &                     \\
	Follow-up time (years)         &             &  & 4.70   & 2.66 -- 5.13        \\
	Hospitalization events (total) & 496        &  &        &                     \\
	Hospitalizations per patient   &             & & 0      & 0 -- 1              \\
	Hospitalization length (days)  &             & & 4      & 3 -- 7              \\ \hline
	\multicolumn{4}{l}{\textit{Variables:}}                                      \\
	Age (years)                    &             &0 & 65     & 59 -- 71            \\
	Female                         & 130 (25.4\%) &0 &        &                     \\
	DLCO                           &              & 5 & 61.0   & 45.8 -- 76.6        \\
	$\text{FEV}_1$                          &              & 0 & 55.6   & 44.9 -- 68.0          \\
	Previous hosp. $=$ 1           & 100 (19.5\%)  & 0 &        &                     \\
	Previous hosp. $\geq$ 2        & 72 (14.1\%)  & 0 &        &                     \\ \hline
	\end{tabular}
	\caption{Descriptive statistics for the whole sample. DLCO = Carbon Monoxide Diffusing Capacity in percentage, $\text{FEV}_1$ = forced expiratory volume in one second in percentage, Previous hosp = hospitalizations during the 2 years prior to the start of follow-up.}
	\label{tab:descriptive}
\end{table}

%%%%%%%%%%%%%%%%%%%%%%%%%%%%%%%%%%%%%%%%%%%%%%%%%%%%%%%%%%%%%%%%%%%%%%%%%%
%%%%%%%%%%%%%%%%%%%%%%%%%%%%%%%%%%%%%%%%%%%%%%%%%%%%%%%%%%%%%%%%%%%%%%%%%%
\section{Dynamic prediction of the risk of death given hospitalization history}
\label{sec:methods}

In this section we obtain the probability of a patient dying between $T$ and $T+w$, given that we observe her hospitalization history up to just before time $T$. The result is based on a joint model for death and hospitalization, which is briefly introduced in Section~\ref{sec:model}. The prediction expression is given in Section~\ref{sec:prediction}. This expression is valid for any model for the hospitalization process. We particularize the expression for a renewal model and thoroughly study how the distribution of hospitalizations impacts the prediction of the risk of death (see Section~\ref{sec:renewal_dependence}).

%%%%%%%%%%%%%%%%%%%%%%%%%%%%%%%%%%%%%%%%%%%%%%%%%%%%%%%%%%%%%%%%%%%%%%%%%%
%%%%%%%%%%%%%%%%%%%%%%%%%%%%%%%%%%%%%%%%%%%%%%%%%%%%%%%%%%%%%%%%%%%%%%%%%%
\subsection{Joint modeling of death and hospitalization}
\label{sec:model}

We briefly introduce the joint model for death, the terminal event, and hospitalization, the recurrent event. A more detailed description can be found in previous work.\citep{liu2004shared,rondeau2007joint} For each patient $i$ in the sample, we observe the following: (i) the death or censoring time $T_i ^d$, (ii) $\delta_i$ an indicator of whether the individual was censored, (iii) $J_i$ the number of hospitalizations prior to censoring or death, (iv) $(T_{ij}^r)_{j=1}^{J_i}$ the times for each hospitalization, and (v) $Z_i$ a vector of baseline covariates. We consider two (possibly overlapping) collections of the variables in $Z_i$, namely $Z_i^d$ and $Z_i^r$, that inform about which variables are considered when modelling the death and hospitalization processes, respectively. 

The death and hospitalization processes are linked by a individual-specific unobserved frailty variable $u_i$. The frailty variable is assumed to be independent of covariates and to have density $g$ supported on $[0, \infty)$. To introduce the model, one must specify $\mathcal{F}_i(t)$ --- the history of the process for patient $i$ up to just before time $t$. The history $\mathcal{F}_i(t)$ consists of knowledge about $T_i^d \geq t$, the frailty variable $u_i$, the covariates $Z_i$, and the history of hospitalizations up to just before time $t$. To ease notation, we denote the history of hospitalizations prior to $t$ by $H_i(t)$. This consists of the number and timing of patient $i$'s hospitalizations before time $t$:
\begin{equation}
	H_i(t) = \left(J_i(t), (T^r_{ij})_{j=1}^{J_i(t)}\right), \text{ where } J_i(t) = \sum_{j=1}^{J_i} \ind(T_{ij}^r < t). 
\end{equation}

The risks for the death and hospitalization processes, $\alpha^d$ and $\alpha^r$ respectively, follow a proportional hazards model with frailty:
\begin{equation}{\label{eq:model}}
	\begin{aligned}
		\alpha^d(t|\mathcal{F}_i(t))&=u_i^\gamma \cdot \exp\left(\beta_d'Z^d_i\right) \cdot \lambda_{0}^d(t), \text{ and } 		\\
		\alpha^r(t|\mathcal{F}_i(t))&=u_i \cdot \exp\left(\beta_r'Z_i^r\right) \cdot \lambda^r(t| H_i(t)). 	
	\end{aligned}
\end{equation}
In the above equations, $\beta_d$ and $\beta_r$ are covariate coefficients, $\gamma$ is a parameter that characterizes the relationship between the processes, $\lambda_0^d$ is the baseline hazard function for the death process, and $\lambda^r$ is the hazard function for the next hospitalization given their history. When $\gamma=0$, the frailty variable does not determine the death risk and the two processes are unrelated. In turn, if $\gamma > 0$, hospitalization and death risk are positively related: a higher risk of death correlates with a higher risk of hospitalization. The opposite happens when $\gamma < 0$. 

The frailty variable $u_i$ is generally assumed to follow a gamma distribution with mean one and variance $\theta$. Estimation of the model is based on parametric maximum likelihood, in case one assumes that the baseline hazard functions are parametric (e.g., Weibull hazards). One could also take a semiparametric approach and approximate the baseline hazards by splines. Estimation is then performed by adding a penalty term to the likelihood that accounts for the complexity of the spline approximation.\citep{rondeau2007joint}

%%%%%%%%%%%%%%%%%%%%%%%%%%%%%%%%%%%%%%%%%%%%%%%%%%%%%%%%%%%%%%%%%%%%%%%%%%
%%%%%%%%%%%%%%%%%%%%%%%%%%%%%%%%%%%%%%%%%%%%%%%%%%%%%%%%%%%%%%%%%%%%%%%%%%
\subsection{Dynamic prediction in joint models}
\label{sec:prediction}

We find an expression for the probability of a patient dying between $T$ and $T+w$, given the observed hospitalization history, as implied by the model in~\eqref{eq:model}. The expression is valid for any model for the hospitalization process, i.e., for any specification of $\lambda^r$. We find that, when hospitalizations follow a renewal model, the predicted risk of death depends on the distribution of hospitalizations throughout the follow-up period.

Consider that we have followed up a patient during $T$ years. We have thus information regarding her hospitalization history during those years --- we know that she was hospitalized exactly in $J$ occasions and that these happened at times $t_1,\dots, t_J$. We refer to this realization of the hospitalization history during follow-up by $h(T) = (J, (t_j)_{j=1}^J)$. That is, $h(T)$ is a specific realization of the random history $H_i(T)$. Also, let $z$ be a realization of the random covariates $Z_i$. The probability of interest is
\begin{equation}
	\mathbb{P}(T, w | z, h(T)) = \prob\left(T_i^d \leq T+w \left| T_i^d \geq T, Z_i=z, H_i(T) = h(T)  \right.\right),
\end{equation}
for $T, w \geq 0$.

We first introduce some definitions that are helpful in obtaining the above probability. For any $t \in [0, T]$, we define the baseline survival function for death as
\begin{equation}
	S_0^d(t) = \exp \left\{ -\int_0^t \lambda_0^d(s)ds \right\}.
\end{equation}
We also introduce the survival function for hospitalizations given their history:
\begin{equation}
	S^r(t|h(t)) = \exp \left\{ -\int_0^t \lambda^r(s|h(s))ds \right\}.
\end{equation}
Note that for each $s\in [0, t]$, $h(s)$ only considers the hospitalizations that happened prior to $s$.  Also, to simplify notation, we denote the proportionality indexes by
\begin{equation}
	C_d = \exp\left\{\beta_d'z^d\right\} \text{ and } C_r = \exp\left\{\beta_r'z^r\right\},
\end{equation}
where $z^d$ and $z^r$ are fixed values for the corresponding covariates in the death and hospitalization processes, respectively. 

We can now present the main result of this section: the risk of death between $T$ and $T+w$ for a patient with covariate values $Z_i=z$ and hospitalization history $H_i(T) = h(T) = (J, (t_j)_{j=1}^J)$. This risk is given by
\begin{equation}{\label{eq:cond_prob_gen}}
	\begin{aligned}
		\mathbb{P}(T, w | z, h(T)) = 
		\frac{\int_0^\infty \left[ S_0^d(T)^{C_d u^\gamma} - S_0^d(T+w)^{C_d u^\gamma} \right]\cdot u^J \cdot S^r(T|h(T))^{C_r u}\cdot g(u)du}{\int_0^\infty S_0^d(T)^{C_d u^\gamma}\cdot u^J \cdot S^r(T|h(T))^{C_r u}\cdot g(u)du}.
	\end{aligned}
\end{equation}
The derivation of the above equation is based on the independence of the death and hospitalization process given frailty (and covariates) and applications of the Bayes rule. We refer to Appendix~\ref{app:prob} for more details.

We highlight that equation~\eqref{eq:cond_prob_gen} is valid for any model for the hazard for the next hospitalization given their history: $\lambda^r$. This hazard may depend on the history of hospitalizations $h(t)$ in various ways. Different models for the hospitalization history will lead to different shapes of $S^r$, the survival function for hospitalizations given their history. Below we discuss the Poisson model, studied by \citet{mauguen2013dynamic}, and the renewal model for the hospitalization process.

\paragraph{Renewal process for hospitalizations} Consider a patient that is hospitalized $J$ times in the follow-up period $[0, T]$. The hospitalization times are $t_1 < \dots < t_J$. In a renewal model (gap time scale) the hazard for the next hospitalization given their history at time $t \in [0, T]$ is
\begin{equation}
	\lambda^r(t|h(t)) = \lambda_0^r\left(t-t_{J(t)}\right),
\end{equation}
where $\lambda_0^r$ is a baseline hazard function (e.g., a Weibull hazard) and $J(t)$ is the number of hospitalizations before time $t$ (cf. with eq.~(1.5) in \citet{cook2007statistical}, where they use counting process notation). That is, the gap time between the last hospitalization and time $t$ is taken into account. Let us now obtain the shape of the survival function for hospitalizations at the end of the follow-up period: $S^r(T|h(T))$. To do so, we partition the follow-up period $[0, T]$ into $J+1$ inter-hospitalization intervals. For convenience, we set
\begin{equation}
	t_0 = 0 \text{ and } t_{J+1} = T.
\end{equation} 
This way, the first inter-hospitalization interval $(t_0, t_1]$ is the time from the beginning of the follow-up to the first hospitalization, while the last interval $(t_J, t_{J+1}]$ is the time from the last hospitalization to the end of the follow-up. We have that
\begin{equation}
	\int_0^T \lambda^r(s|h(s))ds = \int_0^T \lambda_0^r\left(s - t_{ J(s)}\right)ds = \sum_{j=1}^{J+1} \int_{t_{j-1}}^{t_j} \lambda_0^r\left(s-t_{J(s)}\right)ds.
\end{equation}

Now, for $s\in (t_{j-1}, t_j]$, the number of hospitalizations prior to $s$ is $J(s)=j-1$. Therefore
\begin{equation} \label{eq:renewal_hazard}
	\begin{aligned}
		S^r(T|h(T)) &= \exp\left\{ -\int_0^T \lambda^r(s|h(s))ds\right\} = \prod_{j=1}^{J+1} \exp\left\{  - \int_{t_{j-1}}^{t_j} \lambda_0^r\left(s-t_{j-1}\right)ds \right\} \\
		&= \prod_{j=1}^{J+1} \exp\left\{  - \int_{0}^{t_j-t_{j-1}} \lambda_0^r\left(s\right)ds \right\} = \prod_{j=1}^{J+1} S^r_0(t_j-t_{j-1}),
	\end{aligned}
\end{equation}
where $S_0^r(t) = \exp\{-\int_0^t \lambda_0^r(s)ds \}$ is the baseline survival function for hospitalizations. 

The survival function for hospitalization given their history depends on the between hospitalization gap times $t_j - t_{j-1}$. Hence, following equation~\eqref{eq:cond_prob_gen}, the risk of death for two patients, both having $J=2$ hospitalizations, may be different ---it will depend on the distribution of the two hospitalizations. In Section~\ref{sec:renewal_dependence} we discuss which parameters of the model condition the dependence of the risk of death on the distribution of hospitalizations.

\paragraph{Poisson process for hospitalizations}
For the sake of comparison, we recover the results in \citet{mauguen2013dynamic} for the Poisson model (calendar time scale) for hospitalizations. In this model, the hazard for the next hospitalization at time $t \in [0, T]$ is $\lambda^r(t|h(t))=\lambda_0^r(t)$, being $\lambda_0^r$ the baseline hazard for hospitalizations. That is, the hazard for the next hospitalization is independent of the history. Therefore, the survival function for hospitalizations at the end of the follow-up period is $S^r(T|h(T)) = S^r_0(T)$. This solely depends on the follow-up time and not on the hospitalization times. Plugging in this into equation~\eqref{eq:cond_prob_gen} gives equation~(5) in \citet{mauguen2013dynamic}, where only the number of hospitalizations ($J$) matters to predict the risk of death. 

%%%%%%%%%%%%%%%%%%%%%%%%%%%%%%%%%%%%%%%%%%%%%%%%%%%%%%%%%%%%%%%%%%%%%%%%%%
%%%%%%%%%%%%%%%%%%%%%%%%%%%%%%%%%%%%%%%%%%%%%%%%%%%%%%%%%%%%%%%%%%%%%%%%%%
\subsection{Renewal model: relevance of the distribution of hospitalizations}
\label{sec:renewal_dependence}

We have argued that, under the renewal model, the distribution of hospitalization plays a significant role in predicting the risk of death. A natural question to ask then is which pattern of hospitalization times leads to the highest risk of death. Is it when hospitalizations are concentrated or spread out through the follow-up period? In this section we show that the answer depends on two features of the model: the relationship between the two processes ($\gamma$) and the shape of the baseline hazard function for hospitalizations ($\lambda_0^r$). A formal proof of the results is presented in Appendix~\ref{app:distribution_effect}.

Recall that the distribution of hospitalizations determines the risk of death through the value of the survival function of hospitalizations given their history at time $T$. For the renewal model, this is given by
\begin{equation} \label{eq:surv_renewal}
	S^r(T|h(T)) = \prod_{j=1}^{J+1} S^r_0(t_j-t_{j-1}).
\end{equation}
For a fixed number of hospitalizations $J$, the value of the survival function $S^r(T|h(T))$ is determined by the gap times $t_j - t_{j-1}$. We say that hospitalizations are dispersed when they are equiespaced in the $[0, T]$ interval. For instance, a patient with two hospitalizations during the first year of follow-up, one at month 4 and the other at month 8, has dispersed hospitalizations. This translates into three equal gap times of 4 months. Conversely, we regard hospitalizations as concentrated when they are close to each other. For instance, a patient with two hospitalizations at months 8 and 10 has concentrated hospitalizations. This translates into a large gap time of 8 months, followed by two small gap times of 2 months.

Two patients with different distributions of hospitalizations will have a different value of the survival function $S^r(T|h(T))$, even if both have experienced the same number of hospitalizations. To  better understand the relationship between the hospitalization pattern and the risk of death, it is worth focusing on the cumulative hazard function for hospitalizations:
\begin{equation}
	\Lambda^r(T|h(T)) = \int_0^T \lambda^r(s|h(s))ds = \sum_{j=1}^{J+1} \int_{t_{j-1}}^{t_j} \lambda_0^r\left(s-t_{J(s)}\right)ds.
\end{equation} 
The cumulative hazard function is the area under the hazard function of hospitalizations given their history. Note that there is an inverse relationship between the cumulative hazard and the survival function: $S^r(T|h(T)) = \exp\{-\Lambda^r(T|h(T))\}$, cf. equations (1.3) and (1.5) in \citet{aalen2008survival}. Thus, which distribution of hospitalizations leads to larger values of the cumulative hazard (i.e., smaller values of the survival function) at time $T$? This depends on \emph{whether the baseline hazard for hospitalizations is increasing or decreasing}. 

Let us consider two patients under two different scenarios for the baseline hazards. During the first year of follow-up ($T=12$ months), Patient A has dispersed hospitalizations at times $(t_1, t_2)=(4, 8)$. Patient B has concentrated hospitalizations at times $(t_1, t_2)=(8, 10)$. In Scenario 1, the baseline hazard for hospitalizations is increasing: $\lambda_0^r(t)=t$, $t \in [0,T]$. In Scenario 2, the baseline hazard is decreasing: $\lambda_0^r(t)=12-t$, $t \in [0,T]$.

\begin{figure}[h!]
	\centering
	\includegraphics[width=0.95\textwidth]{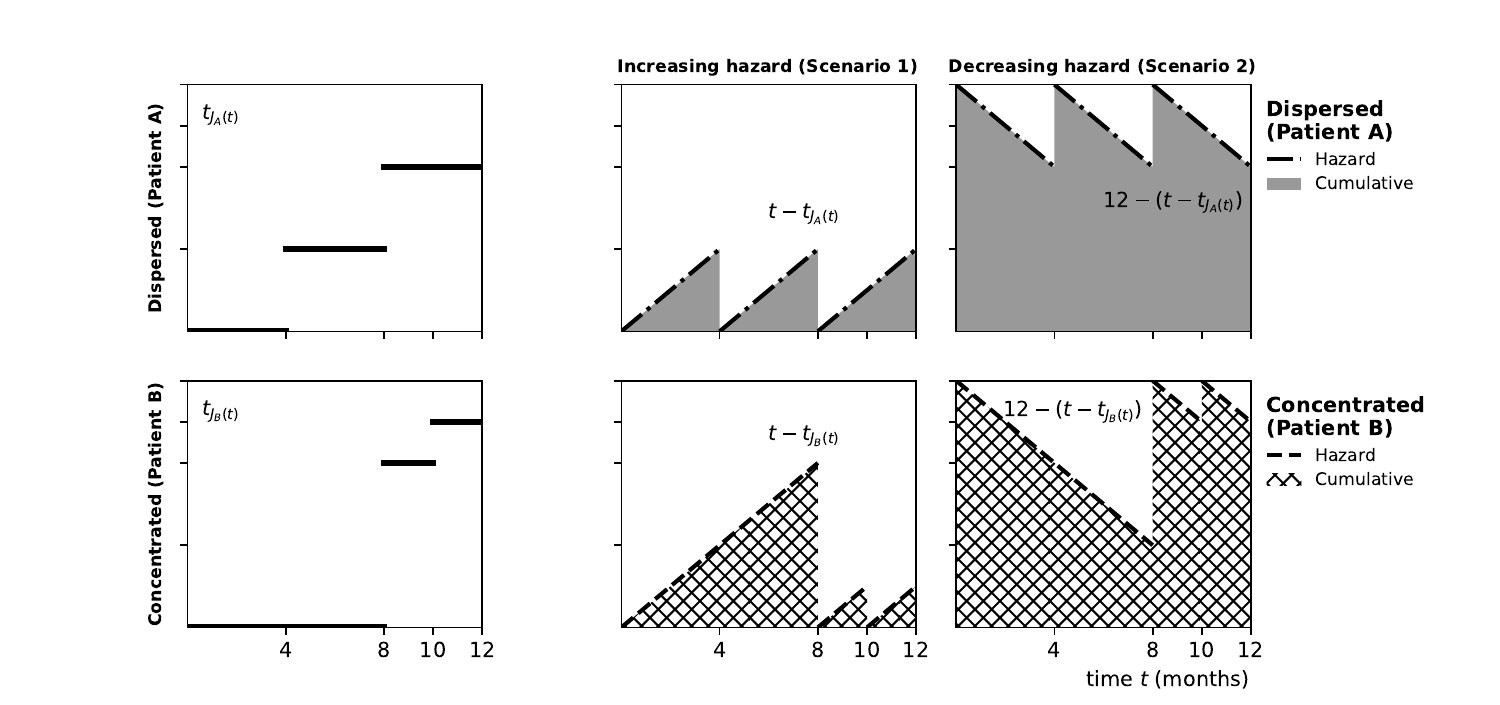}
	\caption{\emph{First column:} Timing of the last hospitalization before time $t$ for Patient A (top row) and Patient B (bottom row).
	\emph{Second and third columns:} Hazard and cumulative hazard for hospitalizations given their history. The second column corresponds to an increasing hazard $\lambda^r_0(t)=t$ (Scenario 1) and the third column to a decreasing hazard $\lambda^r_0(t)=12-t$ (Scenario 2). Top row corresponds to Patient A (dispersed hospitalizations at $t_1=4$ and $t_2=8$) and the bottom row to Patient B (concentrated hospitalizations at $t_1=8$ and $t_2=10$).} 
	\label{fig:cum_hazards}
\end{figure}

The first column of Figure~\ref{fig:cum_hazards} plots the ``timing of the last hospitalization before $t$'' variables $t_{J_A(t)}$ and $t_{J_B(t)}$. For instance, for Patient A (top row), this variable jumps from $0$ to $4$ at time $t=4$, since from that time onwards the last hospitalization happened at time $t_1 = 4$. These are the building blocks to compute the hazard for hospitalizations given their history in the renewal model. 

The second and third columns of Figure~\ref{fig:cum_hazards} plot the hazard for hospitalizations given their history $\lambda^r(t|h(t)) = \lambda_0^r\left(t-t_{J(t)}\right)$, for $t \in [0, T]$. The second column corresponds to an increasing baseline hazard $\lambda_0^r(t) = t$ (Scenario 1) and the third column to a decreasing baseline hazard $\lambda_0^r(t) = 12-t$ (Scenario 2). The top row corresponds to Patient A (dispersed hospitalizations) and the bottom row to Patient B (concentrated hospitalizations). The cumulative hazard (the area under the hazard) is highlighted in all plots. We see that, when the baseline hazard is increasing (Scenario 1), Patient A accumulates less hazard than Patient B. In particular, the values for the cumulative hazard at time $T$ are $24$ and $28$ for Patient A and B, respectively. The opposite happens when the baseline hazard is decreasing (Scenario 2). Patient A accumulates more hazard than Patient B; in particular, the values for the cumulative hazard at time $T$ are $120$ and $116$ for Patient A and B, respectively.

The fact that the relationship between the distribution of hospitalizations and the cumulative hazard (and hence the survival function) depends on the shape of the baseline hazard can be generalized beyond the example (c.f. Proposition~\ref{prop:gap_distribution} in Appendix~\ref{app:distribution_effect}):
\begin{quotation}
	If the baseline hazard for hospitalizations is \emph{increasing}, then \emph{dispersed} hospitalizations lead to larger values of the survival function for hospitalizations at time $T$. Conversely, if the baseline hazard is \emph{decreasing}, then \emph{concentrated} hospitalizations lead to larger values of the survival function.
\end{quotation}

Up to this point, we have studied the impact of the distribution of hospitalizations on the value of the survival function for hospitalizations at time $T$. But what is the effect of the later on the risk of death? Equation~\eqref{eq:cond_prob_gen} characterizes the dependence of the risk of death on the value of the survival function for hospitalizations. We argue that the dependence is determined by the sign of parameter $\gamma$. That is, on whether the death and hospitalization processes are positively or negatively related. 

As we have already done, it is better to study the problem through the lenses of the cumulative harzard for hospitalizations. Consider Scenario 1, where Patient A accumulates less hazard than Patient B. When confronted with this fact, how does the model update the expected value of the frailty variable for each patient? Note that Patient B has been ``exposed'' to a larger cummulative hazard than Patient A, so Patient B was expected to experience more hospitalizations. However, they both experienced the same number of hospitalizations. Thus, Patient B must be less fragile than Patient A. In conclusion, patients with larger values of the cumulative harzard are expected to have smaller frailty values. 

The preceding discussion may be rephrased as ``when comparing two patients with the same number of hospitalizations, patients with larger values of the survival function for hospitalizations ($S^r(t|h(t))$) are expected to have larger frailty values". When $\gamma > 0$, this translates into a higher risk of death. The opposite happens when $\gamma < 0$. We therefore have the following result (c.f. Proposition~\ref{prop:deriv_z} in Appendix~\ref{app:distribution_effect}):
\begin{quotation}
	When the death and hospitalization processes are positively related ($\gamma > 0$), larger values of $S^r(T|h(T))$ imply a higher risk of death. When the relation is negative ($\gamma < 0$), larger values of $S^r(T|h(T))$ imply a lower risk of death. If the processes are unrelated ($\gamma = 0$), the history of hospitalizations has no effect on the risk of death.
\end{quotation}

The two results introduced in this section describe how the distribution of hospitalizations affects the risk of death in the renewal model. Table~\ref{tab:summary} summarizes the main findings (see also Theorem~\ref{thm:charac_renewal} in Appendix~\ref{app:distribution_effect}). For instance, when death and hospitalizations are positively related and the baseline hazard for hospitalizations is decreasing, the risk of death is higher for patients with concentrated hospitalizations.

\begin{table}[!ht]
	\centering
	\resizebox{0.9\textwidth}{!}{%
	\begin{tabular}{cccc}
																		  &            & \multicolumn{2}{c}{Relationship between death and hospitalizations} \\ \cline{3-4} 
																		  &            & Positive ($\gamma>0$)            & Negative ($\gamma<0$)            \\
	\multicolumn{1}{c|}{\multirow{2}{*}{Baseline hazard ($\lambda_0^r$)}} & Increasing & Dispersed                        & Concentrated                     \\
	\multicolumn{1}{c|}{}                                                 & Decreasing & Concentrated                     & Dispersed                       
	\end{tabular}%
	}
	\caption{ The hospitalization pattern that leads to a \emph{higher} risk of death according to the renewal model is shown based on the distribution of hospitalizations and the relationship between death and hospitalization processes. 
		%Summary of the effect of the distribution of hospitalizations in the risk of death, according to the renewal model. The table shows which hospitalization pattern leads to a \emph{higher} risk of death.
	}
	\label{tab:summary}
	\end{table}

	We would like to highlight that the results in Table~\ref{tab:summary} indicate that the renewal model for hospitalizations is flexible enough to provide a wide range of predictions for the risk of death. Depending on the parameters of the model ---$\gamma$ and the shape of the baseline hazard for hospitalizations---, either concentrated or dispersed hospitalizations will be beneficial regarding the risk of death. Practitioners can compare this with their clinical experience. Thus, the insights from this section equip physicians with a tool to qualitatively assess the model's fit.

%%%%%%%%%%%%%%%%%%%%%%%%%%%%%%%%%%%%%%%%%%%%%%%%%%%%%%%%%%%%%%%%%%%%%%%%%%
%%%%%%%%%%%%%%%%%%%%%%%%%%%%%%%%%%%%%%%%%%%%%%%%%%%%%%%%%%%%%%%%%%%%%%%%%%
\section{Risk of death given hospitalization history for COPD patients}
\label{sec:analysis}

We fit a joint model for hospitalization and death of COPD patients. We consider both a Poisson (calendar timescale) model and a renewal (gap timescale) model for hospitalizations. In the Poisson model, time is measured in ``days since inclusion in the study''. In the renewal model, time is measured in ``days since inclusion in the study'' for the first hospitalization and ``days since last hospitalization date'' for the following ones. We note that COPD patients tend to have short hospital stays: median of 4 days (see Table~\ref{tab:descriptive}). Therefore, we keep the timescale as ``days", as opposed to ``days out of hospital".\citep{gonzalez2005sex} Prediction results are more interpretable this way.

For both models, we consider Weibull baseline hazards for death and hospitalization. These are given by $\lambda_0^e(t)=\sigma_e t^{\sigma_e -1 } / b_e^{\sigma_e}$ for $e=d$ (death) and $e=r$ (hospitalizations). Weibull hazards are specifically suited for the problem: they allow us to rapidly understand how the distribution of hospitalizations will affect the risk of death. Indeed, if the shape parameter satisfies $\sigma_r > 1$, the baseline hazard for hospitalizations is strictly increasing. If $\sigma_r < 1$, the baseline hazard is strictly decreasing. We specified a gamma distribution for the frailty variable $u_i$, with density $g(u)=u^{1/\theta-1}\exp(-u/\theta)/(\theta^{1/\theta}\Gamma(1/\theta))$. That is, $u_i$ has mean one and variance $\theta$.

\begin{table}[h!]
	\centering
	\resizebox{.9\textwidth}{!}{
	\begin{tabular}{rlcccc}
									  &                     & \multicolumn{2}{c}{Renewal model}  & \multicolumn{2}{c}{Poisson model}  \\ \cmidrule(rl){3-4}\cmidrule(rl){5-6} 
									  &                     & HR/Estimate & 95\% CI              & HR/Estimate & 95\% CI              \\ \hline
	\multirow{7}{*}{Death}            & Age (years)         & 1.106       & (1.069, 1.143)       & 1.107       & (1.07, 1.146)        \\
									  & Female              & 0.645       & (0.326, 1.277)       & 0.657       & (0.33, 1.309)        \\
									  & DLCO                & 0.961       & (0.947, 0.976)       & 0.960       & (0.946, 0.975)       \\
									  & $\text{FEV}_1$      & 0.982       & (0.966, 0.999)       & 0.982       & (0.965, 0.998)       \\
									  & \multicolumn{5}{l}{\textit{Baseline hazard:}}                                                 \\
									  & Scale ($b_d$)       & 4487.937    & (3297.038, 5678.837) & 4313.069    & (3195.404, 5430.735) \\
									  & Shape ($\sigma_d$)  & 1.723       & (1.41, 2.036)        & 1.748       & (1.431, 2.065)       \\ \hline
	\multirow{9}{*}{Hospitalizations} & Age (years)         & 1.050       & (1.03, 1.071)        & 1.056       & (1.035, 1.078)       \\
									  & Female              & 1.387       & (0.934, 2.059)       & 1.406       & (0.923, 2.141)       \\
									  & DLCO                & 0.985       & (0.976, 0.994)       & 0.983       & (0.973, 0.992)       \\
									  & $\text{FEV}_1$      & 0.962       & (0.951, 0.973)       & 0.959       & (0.947, 0.971)       \\
									  & Prev. hosp $=$ 1    & 1.750       & (1.215, 2.523)       & 1.859       & (1.264, 2.733)       \\
									  & Prev. hosp $\geq$ 2 & 2.283       & (1.547, 3.368)       & 2.632       & (1.745, 3.97)        \\
									  & \multicolumn{5}{l}{\textit{Baseline hazard:}}                                                 \\
									  & Scale ($b_r$)       & 3520.435    & (2485.132, 4555.739) & 2651.037    & (2055.216, 3246.859) \\
									  & Shape ($\sigma_r$)  & 0.857       & (0.789, 0.925)       & 1.142       & (1.042, 1.241)       \\ \hline
	\multirow{2}{*}{Frailty}          & $\gamma$            & 0.730       & (0.359, 1.101)       & 0.715       & (0.381, 1.049)       \\
									  & Variance ($\theta$) & 1.268       & (0.851, 1.685)       & 1.567       & (1.141, 1.993)       \\ \bottomrule
	\end{tabular}%
	}
	\caption{Fit results. Sample: 507 patients, 91 death events, and 473 hospitalization events. \textit{Columns:} HR = Hazard Ratio (computed for covariates), Estimate (computed for scale, shape, and frailty parameters), CI = Confidence Interval. \textit{Covariates:} Age, Female, DLCO = Carbon Monoxide Diffusing Capacity in percentage, $\text{FEV}_1$ = Forced expiratory volume in one second in percentage, Prev. hosp = hospitalizations during the 2 years prior to the start of follow-up. Age, DLCO and $\text{FEV}_1$ expressed as deviations from the baseline patient. \textit{Baseline patient (median):} Age = 65, Sex = Male, DLCO = 61.0, FEV = 55.6, Prev. hosp = 0. \textit{Log-likelihoods:} -4452.081 (renewal) and -4455.676 (Poisson). }
	\label{tab:fit}
\end{table}

Table~\ref{tab:fit} shows the results obtained for the estimated joint model considering both renewal and Poisson specifications for the hospitalization process. We have included age, sex, DLCO, and forced expiratory volume in 1 second in percentage ($\text{FEV}_1$) as common covariates for modelling hospitalization and death processes, and the number of hospitalizations 2 years prior to the first evaluation, for the hospitalization process. This decision was made on the basis of results obtained in previous studies, as well as statistical significance test at 0.05 level. Similar estimates for all the covariates have been obtained in both renewal and Poisson models. In particular in the renewal model, patients with lower values of $\text{FEV}_1$ or DLCO have higher risk for hospitalization and dying: $\text{FEV}_1$ and DLCO Hazard Ratios (HR) are 0.962 and 0.982 for Hospitalization and 0.985 and 0.961 for Death, respectively. Patients face a higher risk as age increases (HR=1.106 and HR=1.050 for Hospitalization and Death, respectively). The results are adjusted by sex, which was not statistically significant (the confidence interval for the HR includes 1). Hospitalization in the two years prior to joining the study were statistically significant for the risk of hospitalization (2 or more HR=2.283). Hospitalization and death risks are positively related ($\gamma=0.730$ and $\gamma=0.715$, in renewal and Poisson models, respectively).

\begin{figure}[!ht]
	\centering
	
	\begin{subfigure}{0.48\linewidth}
		\includegraphics[width=\linewidth]{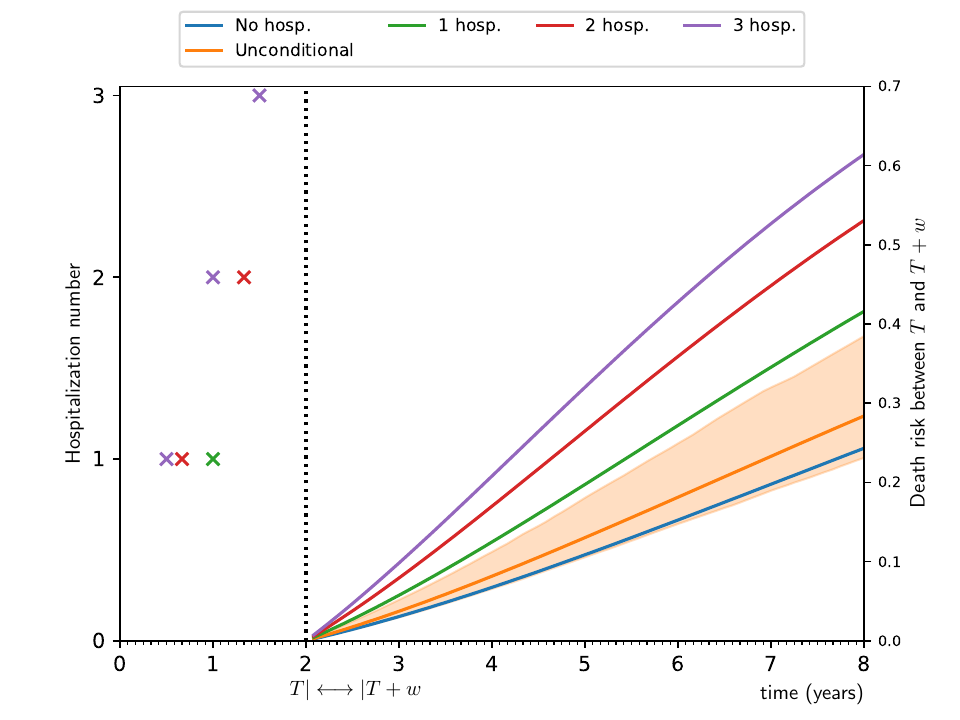}
		\caption{Renewal model. Prediction after 2 years of follow-up.}
	\end{subfigure}\hfill
	\begin{subfigure}{0.48\linewidth}
		\includegraphics[width=\linewidth]{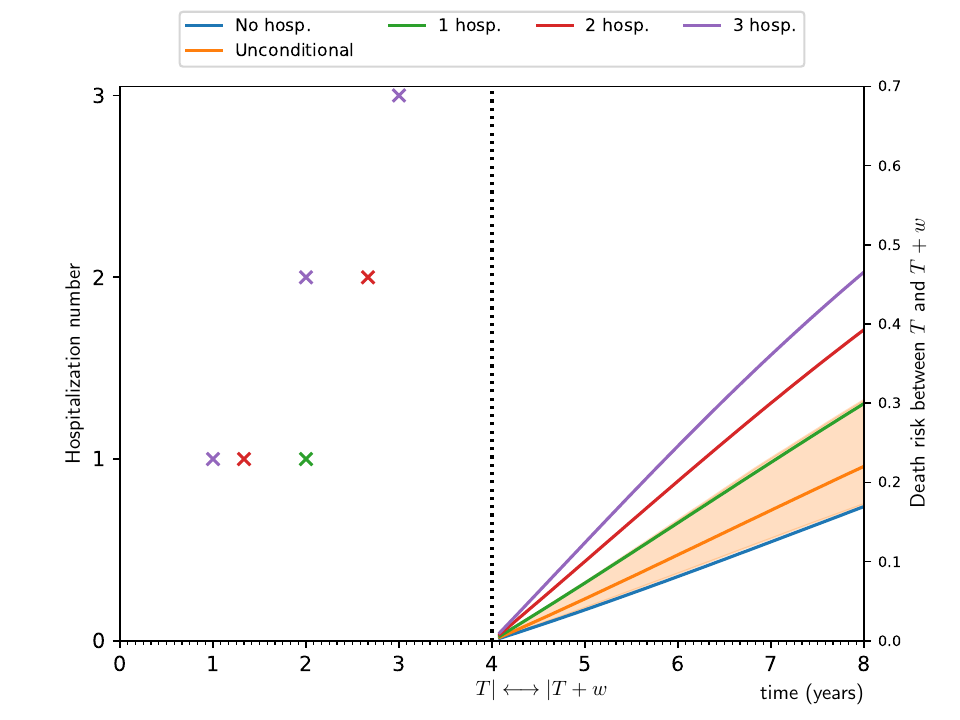}
		\caption{Renewal model. Prediction after 4 years of follow-up.}
	\end{subfigure}

	\begin{subfigure}{0.48\linewidth}
		\includegraphics[width=\linewidth]{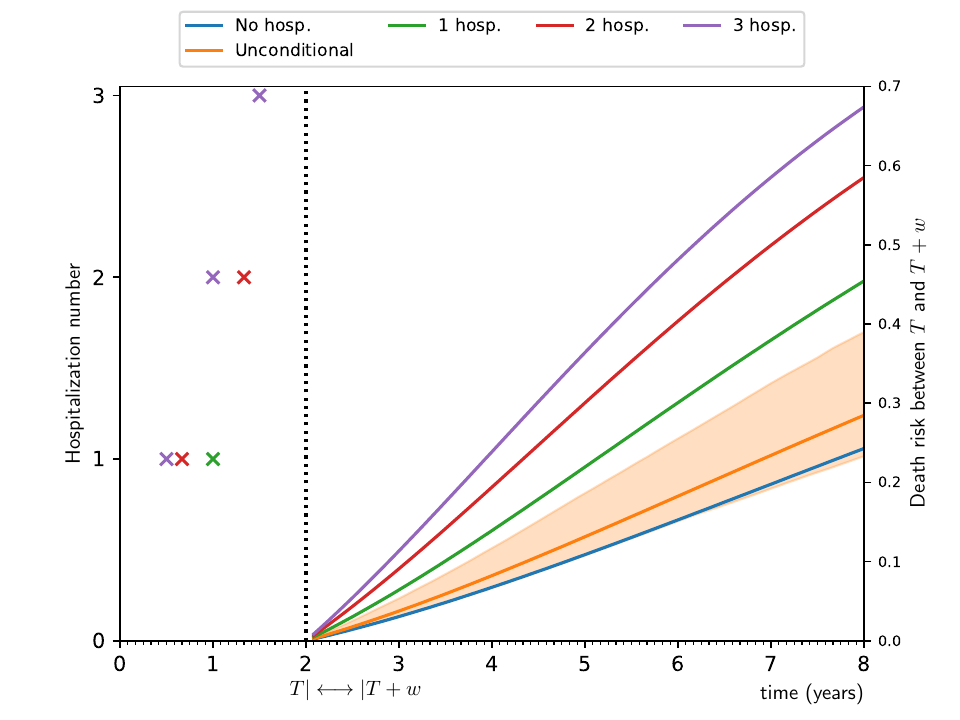}
		\caption{Poisson model. Prediction after 2 years of follow-up.}
	\end{subfigure}\hfill
	\begin{subfigure}{0.48\linewidth}
		\includegraphics[width=\linewidth]{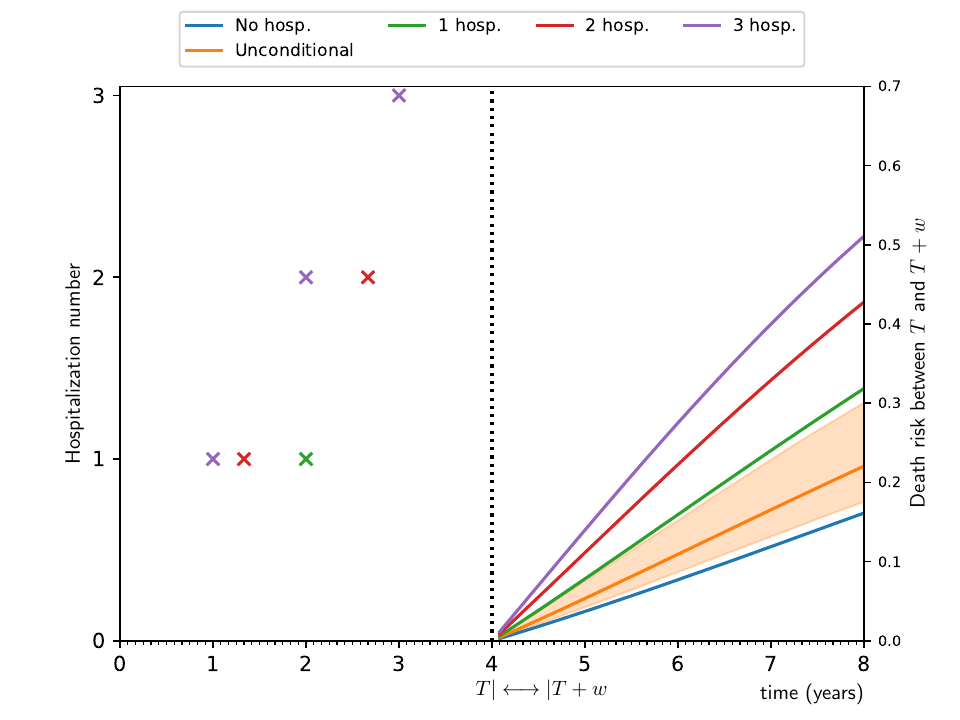}
		\caption{Poisson model. Prediction after 4 years of follow-up.}
	\end{subfigure}

	\caption{Prediction of the risk of death given the hospitalization history at different follow-up times. Results for the median patient (65 years-old, male, DLCO of 61.0, $\text{FEV}_1$ of 55.6, and with no hospitalizations prior to follow-up) differing on the number of hospitalizations (0 to 3). ``Unconditional'' means not conditioning on the hospitalization history. Shaded area gives the 90\% confidence interval for ``Unconditional".}
	\label{fig:predict_multihosp}
\end{figure}

Figure~\ref{fig:predict_multihosp} shows predictions of the risk of death given the history of hospitalizations for four different patients. All of them resemble the median patient (65 years-old, male, DLCO of 61.0, $\text{FEV}_1$ of 55.6, and with no hospitalizations prior to follow-up) but experienced a different number of hospitalizations during follow-up: from zero to three. The figure also shows the ``Unconditional'' risk of death, i.e., not conditioning on the hospitalization history.\citep{mauguen2013dynamic} Predictions were made at the $T=2$ and $T=4$ year marks after follow-up began. Each panel displays the risk of death between $T$ and $T+w$. For instance, the risk of death of the median patient who experiences one hospitalization in the first two years of follow-up, as predicted by the renewal model, is given by the green line in Panel (a). Its value at $T+w=5$ is $0.2195$. This means that the risk of death between the second and fifth year, conditional on surviving up to the second, is 21.95\%. 

Several conclusions can be obtained from these results. First, and as it was previously mentioned, conditioning on the number os hospitalizations has an impact on the predicted risk of death, with increasing number of hospitalizations increasing the risk of death. Second, for predictions shortly after follow-up begins, ``Unconditional'' is close to ``No hospitalizations''. This means that no hospitalizations are generally expected in short follow-up periods. However, for larger follow-up periods (see Panels (b) and (d)), ``Unconditional'' gets closer to ``1 hospitalization'' which is consistent with the observed data where the mean number of hospitalizations per patient is 0.9329. 

Third, the predicted risk is smaller under the renewal model than the one under the Poisson model (see for example Panel (a) and Panel (c)). The difference is remarkable for patients experiencing a large number of hospitalizations: two or three, corresponding to the red and purple lines, respectively. In the Poisson model, hospitalizations are independent events. Therefore, a large number of hospitalizations can only be seen as a sign of a large frailty. In the renewal model, the risk of hospitalizations is not only explained by the frailty variable, but by the time since the last hospitalization. Then, if a patient just had an hospitalization, is expected that the risk of another one is high, even if the patient's frailty is not. This can also be seen in the variance of the frailty variable, which is larger for the Poisson model (see the last row in Table~\ref{tab:fit}).

Table~\ref{tab:HRS_diffhosp} shows the hazard ratios for a follow-up time of $T$. In particular, we can observe the HR for a given number of hospitalizations with respect to not having had any hospitalization up to that time. As can be seen, whatever the follow-up time and for both specifications, increasing the number of hospitalizations during follow-up increases the risk of mortality (for a 2-year follow-up, the HR under de renewal specification are 1.898 and 4.435 for 1 and 5 hospitalizations, respectively). This result contrasts with the result obtained in \citet{suissa2012long}, where an attenuation bias is observed since the HR stabilizes after 3 hospitalizations. In addition, in line with the results shown in Figure~\ref{fig:predict_multihosp}, HRs are higher for the Poisson model than for the renewal model.

\begin{table}[ht]
    \centering
	\begin{tabular}{rlccccc}
		\hline
								 &         & \multicolumn{5}{c}{Follow-up time in years ($T$)} \\ \hline
								 &         & 1        & 2        & 4       & 6       & 8       \\ \hline
		\multirow{2}{*}{1 hosp.} & Renewal & 1.906    & 1.898    & 1.895   & 1.901   & 1.91    \\
								 & Poisson & 2.118    & 2.130    & 2.150   & 2.168   & 2.186   \\ \hline
		\multirow{2}{*}{2 hosp.} & Renewal & 2.659    & 2.646    & 2.638   & 2.648   & 2.669   \\
								 & Poisson & 3.01     & 3.054    & 3.106   & 3.147   & 3.187   \\ \hline
		\multirow{2}{*}{3 hosp.} & Renewal & 3.317    & 3.308    & 3.299   & 3.313   & 3.343   \\
								 & Poisson & 3.745    & 3.849    & 3.959   & 4.025   & 4.088   \\ \hline
		\multirow{2}{*}{4 hosp.} & Renewal & 3.892    & 3.903    & 3.904   & 3.923   & 3.962   \\
								 & Poisson & 4.334    & 4.525    & 4.736   & 4.838   & 4.923   \\ \hline
		\multirow{2}{*}{5 hosp.} & Renewal & 4.383    & 4.435    & 4.467   & 4.493   & 4.54    \\
								 & Poisson & 4.793    & 5.084    & 5.444   & 5.601   & 5.709   \\ \bottomrule
		\end{tabular}
    \caption{Hazard Ratios at different follow-up times for patients with different number of hospitalizations. For two hospitalization histories $h(T)$ and $h'(T)$, the Hazard Ratio is approximated by $\mathbb{P}(T, w | z, h(T))/\mathbb{P}(T, w | z, h'(T))$ for a short window ($w=7$ days). The reference (denominator) has no hospitalizations prior to $T$. Results for the median patient (65 years-old, male, DLCO of 61.0, $\text{FEV}_1$ of 55.6, and with no hospitalizations prior to follow-up).}
    \label{tab:HRS_diffhosp}
\end{table}

\subsection{Dependence on the distribution of hospitalizations}

The key result from Section~\ref{sec:renewal_dependence} is that, in a renewal model for the hospitalization process, the risk of death given hospitalizations depends on the distribution of the later. The dependence is characterized by the relationship between the processes (parameter $\gamma$) and the shape of the baseline hazard for hospitalizations (which, in the Weibull case, corresponds to the parameter $\sigma_r$). The estimates of these parameters are (see Table~\ref{tab:fit})
\begin{equation}
	\hat\gamma = 0.730 \text{ and } \hat{\sigma}_r = 0.857.
\end{equation}
That is, the death and hospitalization processes are positively related and the baseline hazard for hospitalizations is decreasing in time. According to our results, this means that the risk of death is lowest for dispersed hospitalizations (see Section~\ref{sec:renewal_dependence}).

\begin{figure}[!ht]
	\centering
	
	\begin{subfigure}{0.48\linewidth}
		\includegraphics[width=\linewidth]{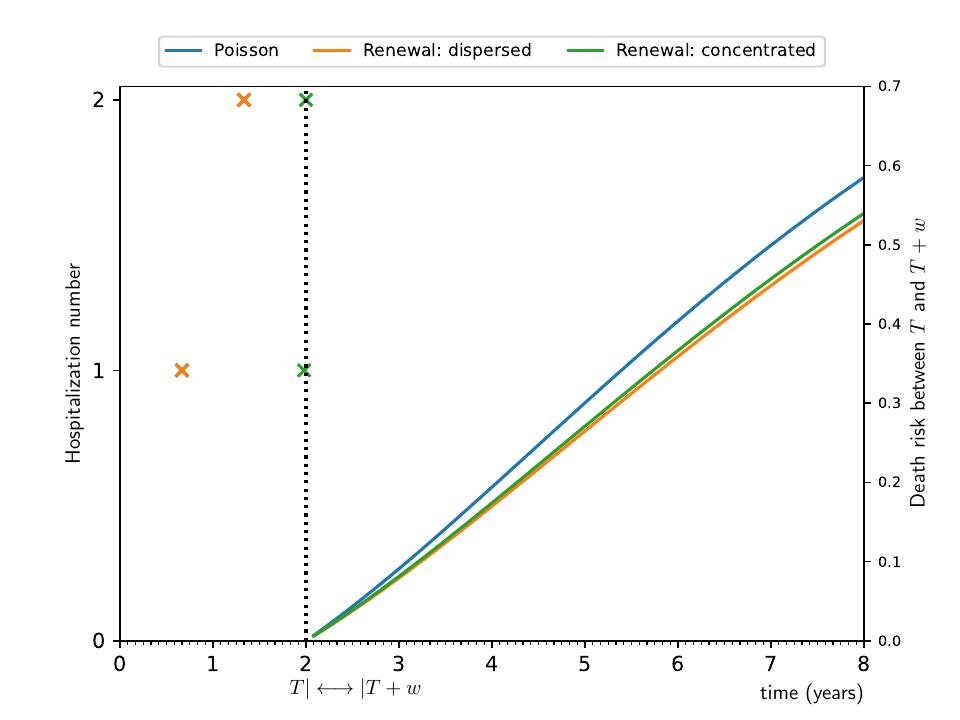} 
		\caption{2 hospitalizations. Prediction after 2 years of follow-up.}
	\end{subfigure}\hfill
	\begin{subfigure}{0.48\linewidth}
		\includegraphics[width=\linewidth]{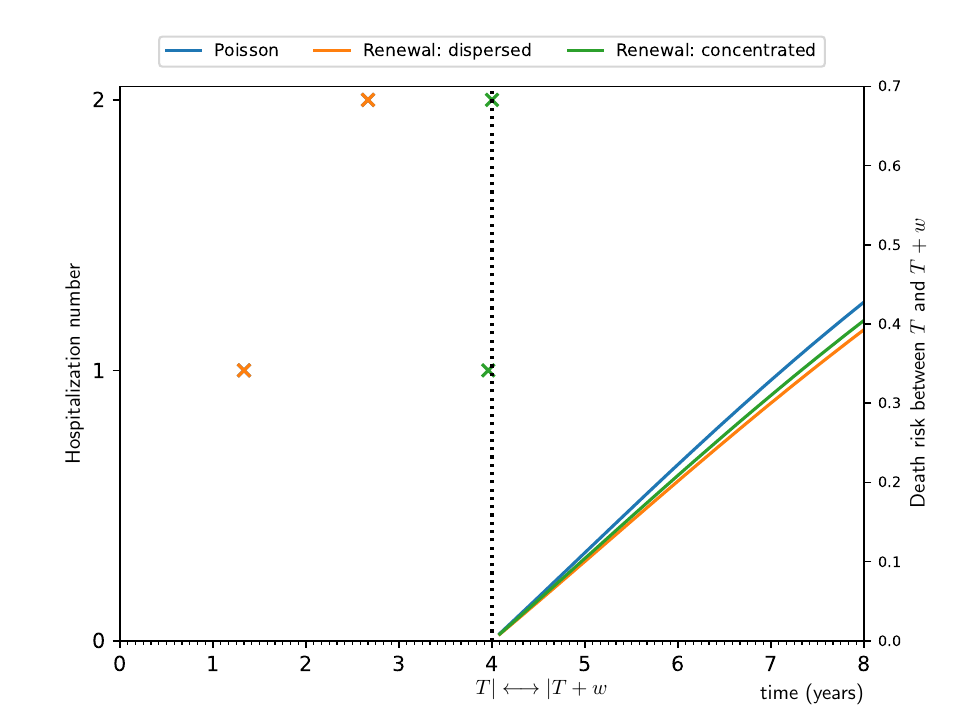}
		\caption{2 hospitalizations. Prediction after 4 years of follow-up.}
	\end{subfigure}
	
	\begin{subfigure}{0.48\linewidth}
		\includegraphics[width=\linewidth]{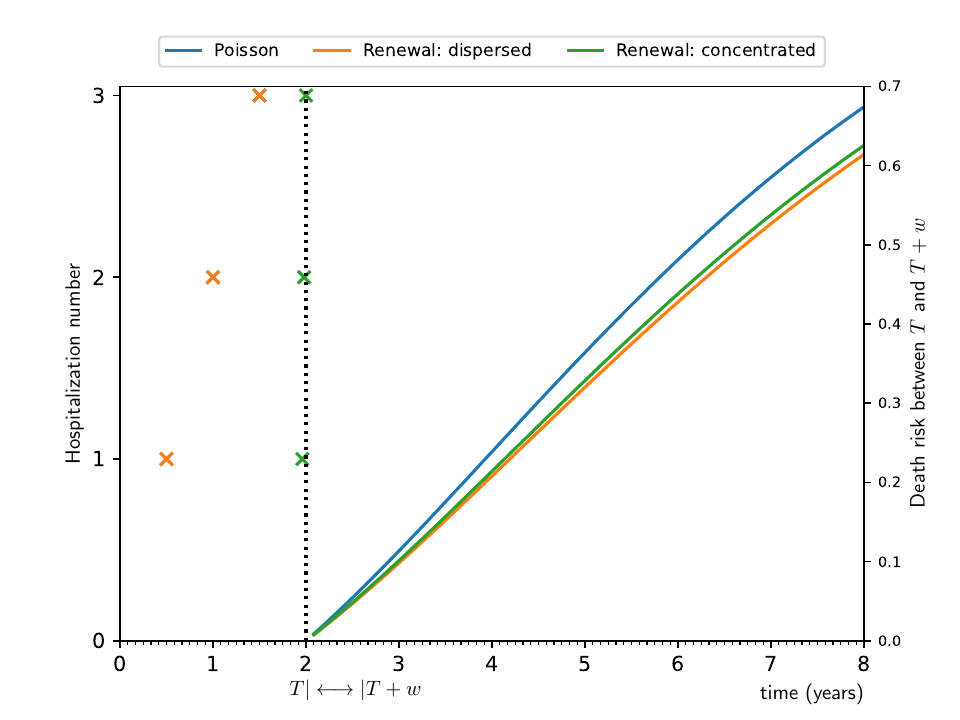}
		\caption{3 hospitalizations. Prediction after 2 years of follow-up.}
	\end{subfigure}\hfill
	\begin{subfigure}{0.48\linewidth}
		\includegraphics[width=\linewidth]{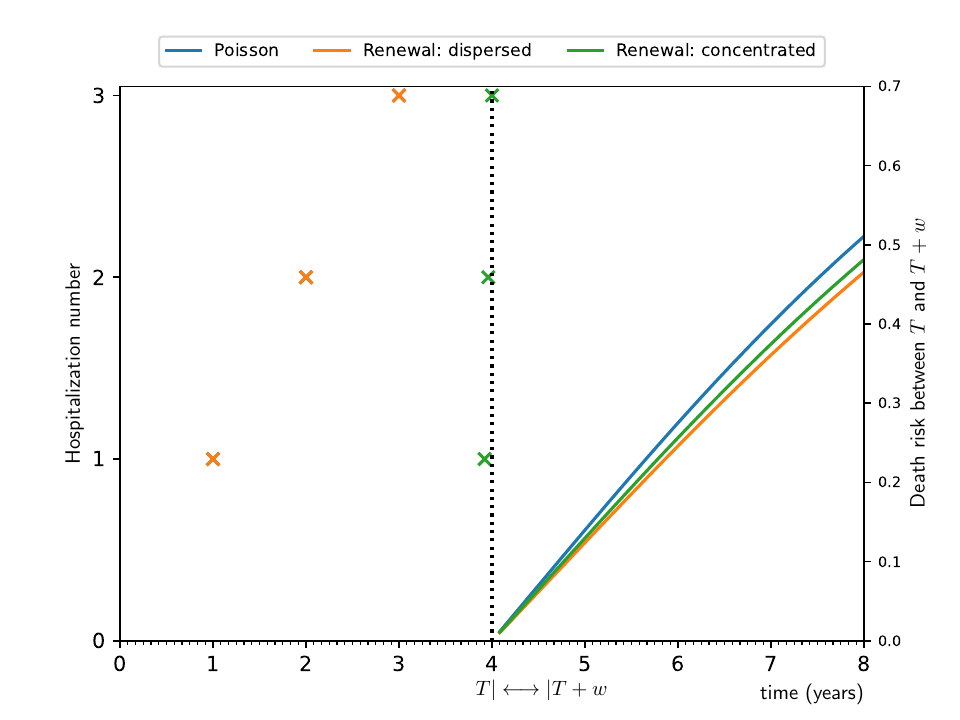}
		\caption{3 hospitalizations. Prediction after 4 years of follow-up.}
	\end{subfigure}

	\caption{Prediction of the risk of death given the hospitalization history at different follow-up times. Results for the median patient (65 years-old, male, DLCO of 61.0, $\text{FEV}_1$ of 55.6, and with no hospitalizations prior to follow-up) differing on the timing of the hospitalizations. Letting $T$ denote the follow-up time, dispersed hospitalizations occur at points $(T/3, 2T/3)$ and $(T/4, T/2, 3T/4)$ for the cases of 2 and 3 hospitalizations, respectively. Concentrated hospitalizations occur at points $(99T/100, T)$ and $(98T/100, 99T/100, T)$ for the cases of 2 and 3 hospitalizations, respectively. For the Poisson model, the distribution of hospitalizations is irrelevant (the number matches each case).}
	\label{fig:predict_dist}
\end{figure}

Figure \ref{fig:predict_dist} shows the prediction of the risk of death given the hospitalization history for four patients with median covariate values. The patients experienced a different number of hospitalizations (2 or 3) during follow-up. Moreover, the distribution of the hospitalizations is dispersed (in orange) or concentrated (in green). As can be observed in all the panels, different predicted risks are estimated (under the renewal model) when hospitalizations occur in a dispersed manner (in orange), or all of them occur in a concentrated pattern (in green), with a higher risk of death for those patients with concentrated hospitalizations. 

%The differences in prediction are sizable. Consider the probability of death between the fourth and eighth years for the median patient with three hospitalizations (Panel (d)). The risk ratio between the patient with dispersed hospitalizations and the patient with concentrated hospitalizations is 1.039. To put it in perspective, this risk ratio is equivalent to an increase of eight months and a half in the patient's age.

The differences in terms of risk of death between concentrated and dispersed hospitalizations are sizeable. Table~\ref{tab:HRS_distribution} shows the Hazard Ratios at different follow-up times for patients with different distribution of hospitalizations (renewal model)  conditioned on all the covariates of the model (i.e. age, sex, DLCO, FEV1). Note that even after conditioning for these clinical and sociodemographic factors, the distribution of the hospitalizations is an important risk factor of death. For instance, at the fourth year, the HR of 3 concentrated hospitalizations versus 3 dispersed hospitalizations (HR=1.050) is equivalent to the HR associated with being 6 months older, holding the remaining clinical characteristics fixed. In addition, for a given number of hospitalizations, the risk of concentrated versus dispersed hospitalizations increases as the follow-up time increases. 

\begin{table}[ht]
    \centering
	\begin{tabular}{rccccc}
			& \multicolumn{5}{c}{Follow-up time in years ($T$)} \\ \cline{2-6} 
			& 1        & 2        & 4       & 6       & 8       \\ \hline
	2 hosp. & 1.017    & 1.027    & 1.041   & 1.048   & 1.052   \\
	3 hosp. & 1.019    & 1.033    & 1.050   & 1.059   & 1.065   \\
	4 hosp. & 1.02     & 1.035    & 1.056   & 1.068   & 1.074   \\
	5 hosp. & 1.019    & 1.035    & 1.06    & 1.074   & 1.081   \\ \hline
	\end{tabular}
    \caption{Hazard Ratios at different follow-up times for patients with different distribution of hospitalizations (renewal model). For two hospitalization histories $h(T)$ and $h'(T)$, the Hazard Ratio is approximated by $\mathbb{P}(T, w | z, h(T))/\mathbb{P}(T, w | z, h'(T))$ for a short window ($w=7$ days). Fixed the number of hospitalizations, we compute the Hazard Ratio of concentrated hospitalizations over dispersed hospitalizations. Results for the median patient (65 years-old, male, DLCO of 61.0, $\text{FEV}_1$ of 55.6, and with no hospitalizations prior to follow-up).}
    \label{tab:HRS_distribution}
\end{table}

Finally, we propose to statistically test whether the distribution of hospitalization determines the shape of the risk of death. Note that, for a Weibull hazard, the distribution of hospitalizations is irrelevant if either $\gamma = 0$ or $\sigma_r = 1$. That is, the null hypothesis of no relevance of the distribution of hospitalizations is $\operatorname{H}_0\colon \gamma=0 \text{ or } \sigma_r=1$. We construct a Wald test for the equivalent hypothesis $\operatorname{H}_0\colon \gamma \cdot (\sigma_r-1) = 0 $ versus $\operatorname{H}_1\colon \gamma \cdot (\sigma_r-1) \neq 0 $. Following \citet[p.~433]{shao2003mathematical}, the test statistic is
\begin{equation}
	\hat{W} \equiv  \frac{R(\hat{\gamma}, \hat{\sigma}_r)^2}{\nabla R(\hat{\gamma}, \hat{\sigma}_r)'\hat{V}\nabla R(\hat{\gamma}, \hat{\sigma}_r)},
\end{equation}
where $ R(\gamma, \sigma_r) = \gamma \cdot (\sigma_r-1) $, $\nabla R(\gamma, \sigma_r) = (\sigma_r-1, \gamma)'$ is the gradient of $R$, $(\hat{\gamma}, \hat{\sigma}_r)$ is the estimator of the parameters, and $\hat{V}$ is an estimator of the asymptotic covariance matrix of $(\hat{\gamma}, \hat{\sigma}_r)$. Under $\operatorname{H}_0$, it holds that $\hat{W} \xrightarrow{P} \chi_1^2$. We get that, for our fit of the renewal model, $\hat{W} = 6.3$ (p-value of $0.012$). Thus, we reject the null hypothesis of no effect of the distribution of hospitalizations on the risk of death.

%%%%%%%%%%%%%%%%%%%%%%%%%%%%%%%%%%%%%%%%%%%%%%%%%%%%%%%%%%%%%%%%%%%%%%%%%%
%%%%%%%%%%%%%%%%%%%%%%%%%%%%%%%%%%%%%%%%%%%%%%%%%%%%%%%%%%%%%%%%%%%%%%%%%%
\section{Discussion}
\label{sec:discussion}

We have developed a general dynamic prediction framework for the risk of a terminal event (death) given the recurrent event (hospitalization) history in a joint model setting. The results allow for any model for the hospitalization process. Prediction results in this setting, solely for a Poisson process for the recurrent event, where obtained in previous work.\citep{mauguen2013dynamic} For prediction of a terminal event given the history of a time-dependent marker, see \citet{rizopoulos2012joint} and references therein.

We have studied how the distribution of hospitalizations throughout the follow-up period determines the risk of death when hospitalizations follow-up a renewal process. In contrast to the Poisson case, where solely the number of hospitalization during follow-up matters, we have found that the risk of death depends on the gap times between hospitalizations. The dependence between the risk of death and the distribution of hospitalizations is characterized by two features: whether the two processes are positively or negatively related (parameter $\gamma$) and whether the baseline hazard for hospitalizations is increasing or decreasing. 

We have focused in a Weibull model for the baseline hazards. The hazard of this model is monotone, with a single parameter determining whether it is increasing or decreasing. This eases the characterization of the dependence between the risk of death and the distribution of hospitalizations, as it is reduced to two parameters. Nevertheless, our prediction framework allows for other specifications of the baseline hazard. Spline approximations as in \citet{rondeau2007joint} may also be considered when fitting the model. In those cases, if the baseline hazard for hospitalizations is non-monotone, the dependence of the risk of death and the distribution of hospitalizations should be studied for each case ---i.e., for each patient and prediction window.

Time-dependent external covariates could be included in the hazard for any of the two processes. Note that time-dependent covariates are usually assumed to be constant in between hospitalizations (see p.~65 in \citet{cook2007statistical}). Under that assumption, one could generalize our results to allow for external time-dependent covariates in the hospitalization process. Regarding time-dependent covariates in the death process, one faces an additional obstacle: the value of the covariates must be known in the time interval $[T, T+w]$. This information is not generally available at the prediction point $T$. So, to predict with external time-dependent covariates in the death process, the researcher must be willing to assume that the value of these covariates remains unchanged for the whole time interval $[T, T+w]$.

We applied our methodology to a dataset of patients with COPD. We found that the risk of death is generally higher when the recurrent hospitalization process is modeled using the Poisson specification. We believe that this is a consequence of hospitalizations being independent in the Poisson model. We see that, in both models, the number of hospitalizations greatly impacts the risk of death. Furthermore, in the renewal model, the distribution of hospitalizations is an important risk factor. 

There are different avenues for future work. Our results correspond to Setting~1 in \citet{mauguen2013dynamic} ---we believe that our results could be readily extended to cover Setting~2. Moreover, to increase prediction accuracy, it may be of interest to consider the evolution of different biomarkers alongside with the history of hospitalizations. The present paper has solely considered a joint model for the death and hospitalization process. Nevertheless, since biomarker data is often available for chronic patients, it is appealing to include it in the model. This requires proposing a joint model for a terminal, a recurrent, and a longitudinal outcome, as in \citet{krol2018multivariate}. Prediction of the risk of the terminal event given both the history of recurrent events and the biomarker may be a promising avenue for future research.

 Additionally, our study of the renewal model for hospitalizations has disclosed some of its limitations. In the renewal model, the distribution of hospitalizations affects the risk of death through the gap times between hospitalizations. Say, for instance, that we have followed up two patients for one year. The first was hopitalized at months 1 and 2. The second was hospitalized at months 10 and 11. These two patients have the same gap times between hospitalizations: a large gap of 10 months and two short gaps of 1 month. The model thus predicts the same risk of death for both patients. 
 
 In view of this result, it seems interesting to model hospitalizations as a process that contains a Poisson and a renewal component\citep{ng1997modeling} or as a self-exciting process.\citep{kopperschmidt2013statistical} This way, one may be able to incorporate differences between hospitalizations at the beginning and at the end of the follow-up period. Our prediction result in equation~\eqref{eq:cond_prob_gen} is valid for arbitrary models of the hospitalization process and is therefore amenable for such extension. The study of a model with Poisson and renewal components and the characterization of the risk of death given different distributions of hospitalizations is left for future work. 

Finally, it is worth mentioning that the software developed to implement the methodological proposal presented in this paper is available on github: \url{https://github.com/telmoperiz/frailtyPredict}.

\section*{Acknowledgments}

This work was financially supported in part by grants from the Departamento de Educaci\'on, Pol\'itica Lingü\'istica y Cultura del Gobierno Vasco IT1456-22, by the Ministry of Science and Innovation through BCAM Severo Ochoa accreditation CEX2021-001142-S / MICIN / AEI / 10.13039/501100011033, by the Basque Government through the BERC 2022-2025 program, the BMTF ``Mathematical Modeling Applied to Health" Project and the Network for Research on Chronicity, Primary Care, and Health Promotion (RICAPPS) and the Instituto de Salud Carlos III (PI13/02352).

%%%%%%%%%%%%%%%%%%%%%%%%%%%%%%%%%%%%%%%%%%%%%%%%%%%%%%%%%%%%%%%%%%%%%%%%%%
%%%%%%%%%%%%%%%%%%%%%%%%%%%%%%%%%%%%%%%%%%%%%%%%%%%%%%%%%%%%%%%%%%%%%%%%%%
\cleardoublepage
\begin{appendices}
	
\section{Risk of death given the hospitalization history}
\label{app:prob}

As in \citet{mauguen2013dynamic}, the derivation in this Appendix is based on the independence of the death and hospitalization process given frailty (and covariates) and applications of the Bayes rule. Additionally, Theorem~2.1 in \citet{cook2007statistical} is used to compute the probability density of a given hospitalization history. This result is valid for arbitrary intensity-based models for the hospitalization process.

The probability of interest is
\begin{equation}
	\mathbb{P}(T, w | z, h(T)) = \prob\left(T_i^d \leq T+w \left| T_i^d \geq T, Z_i=z, H_i(T) = h(T)  \right.\right),
\end{equation}
for $T, w \geq 0$ and hospitalization history $h(T) = (J, (t_j)_{j=1}^J)$. As in \citet{mauguen2013dynamic}, conditioning on the frailty variable one gets
\begin{equation}
	\begin{aligned}
		\mathbb{P}(T, w | z, h(T)) = \\ 
		\int_0^\infty \prob(T_i^d\leq T+w|T_i^d>T, Z_i=z, H_i(T) = h(T), u_i=u)\cdot g(u|T^d_i>T, Z_i=z, H_i(T) = h(T))du.
	\end{aligned}
\end{equation}
We deal with the terms inside the integral separately.

First, using conditional independence of the two processes given frailty and covariates:
 \begin{equation}
 	\begin{aligned}
 		\prob(T_i^d\leq T+w|T_i^d>T, Z_i, H_i(T), u_i)&=\frac{\prob(T<T_i^d\leq T+w|Z_i, H_i(T), u_i)}{\prob(T_i^d >T |Z_i, H_i(T), u_i)} \\
 		&=\frac{\prob(T<T_i^d\leq T+w|Z_i,  u_i)}{\prob(T_i^d >T|Z_i, u_i)} \\
 		&=\frac{\prob(T_i^d >T|Z_i, u_i) - \prob(T_i^d >T+w|Z_i, u_i)}{\prob(T_i^d >T|Z_i, u_i)}.
 	\end{aligned}
 \end{equation}
Furthermore, the survival function for death given covariates and the frailty term takes the following shape:
\begin{equation}
	\prob(T_i^d >T|Z_i, u_i) = \exp\left\{ - \int_0^T \alpha^d(s|\mathcal{F}_i(s))ds \right\}.
\end{equation}
This can be obtained by applying Theorem~2.1 in \citet{cook2007statistical} to the ``0 death events before $T$'' case, noting that (i) ``0 death events before $T$'' is equivalent to $T_i^d  > T$ and (ii) $\mathcal{F}_i(0)$ is the $\sigma$-algebra generated by $(Z_i, u_i)$. By the model in \eqref{eq:model}, the above equation becomes
\begin{equation} \label{eq:cond_surv_death}
	\begin{aligned}
		\prob(T_i^d >T|Z_i, u_i) &= \exp\left\{ u_i^\gamma \cdot  \exp\left(\beta_d'Z^d_i\right) \cdot  \left( - \int_0^T \lambda_{0}^d(s) ds \right) \right\} \\
		&= \exp\left\{ u_i^\gamma \cdot  \exp\left(\beta_d'Z^d_i\right) \cdot  \log S_0^d(T) \right\} \\
		&= \exp\left\{  \log\left( S_0^d(T)^{C_{di}u_i^\gamma} \right)\right\} = S_0^d(T)^{C_{di}u_i^\gamma},		
	\end{aligned}
\end{equation}
where $C_{di} = \exp\left(\beta_d'Z^d_i\right)$. Thus, the first part of the integral is
\begin{equation}
	\prob(T_i^d\leq T+w|T_i^d>T, Z_i, H_i(T), u_i) = \frac{S_0^d(T)^{C_{di}u_i^\gamma} - S_0^d(T+w)^{C_{di}u_i^\gamma}}{S_0^d(T)^{C_{di}u_i^\gamma}}.
\end{equation}

To compute the second part of the integral, we use Bayes' rule to write that the conditional density of the frailty variable as
\begin{equation}{\label{eq:conditional_density}}
	\begin{aligned}
	g(u|T^d_i>T, Z_i=z, H_i(T) = h(T))= \\
	\frac{\prob(T_i^d>T, H_i(T) = h(T)|Z_i=z, u_i=u)g(u)}{\int_0^\infty \prob(T_i^d>T, H_i(T) = h(T)|Z_i=z, u_i=u)g(u)du},
	\end{aligned}
\end{equation}
where we rely on $Z_i$ and $u_i$ being independent. Moreover, by conditional independence of the two processes given frailty and covariates:
\begin{equation}
			 \prob(T_i^d>T, H_i(T) = h(T)|Z_i, u_i)= \prob(T_i^d>T |Z_i, u_i)\cdot \prob(H_i(T) = h(T)|Z_i,u_i).
\end{equation}
The first term in the product follows from equation~\eqref{eq:cond_surv_death}. We apply Theorem~2.1 in \citet{cook2007statistical} to obtain the second term:
\begin{equation}
	\prob(H_i(T) = h(T)|Z_i,u_i) = \left(\prod_{j=1}^{J} \alpha^r(t_j|\mathcal{F}_i(t_j))\right) \cdot \exp\left\{-\int_0^T \alpha^r(s|\mathcal{F}_i(s))ds\right\}.
\end{equation}
Then, by the model in \eqref{eq:model},
\begin{equation}
	\begin{aligned}
		\prod_{j=1}^{J} \alpha^r(t_j|\mathcal{F}_i(t_j)) &= u_i^J  C_{ri}^J \prod_{j=1}^J \lambda^r(t_j|H_i(t_j)) \text{ and } \\
		\exp\left\{-\int_0^T \alpha^r(s|\mathcal{F}_i(s))ds\right\} &= S^r(T|h(T))^{C_{ri}u_i},
	\end{aligned}
\end{equation}
where $C_{ri} = \exp\left(\beta_r'Z^r_i\right)$. The result in the second row follows \textit{mutatis mutandis} from equation~\eqref{eq:cond_surv_death}.
 
 We can now plug in the above results into equation~\eqref{eq:conditional_density}. A key aspect of the model is that, since $ C_{ri}^J \prod_{j=1}^J \lambda^r(t_j|H_i(t_j))$ does not depend on $u_i$, it cancels out. Thus
 \begin{equation}
 	g(u|T^d_i>T, Z_i=z, H_i(T) = h(T))=\frac{S_0^d(T)^{C_{d}u^\gamma}\cdot u^J\cdot S^r(T|h(T))^{C_r u} \cdot g(u)}{\int_0^\infty S_0^d(T)^{C_{d}u^\gamma}\cdot u^J\cdot S^r(T|h(T))^{C_r u} \cdot g(u)du}.
 \end{equation}

We conclude by putting all the results together to obtain equation~\eqref{eq:cond_prob_gen}:
\begin{equation}
	\begin{aligned}
		\mathbb{P}(T, w | z, h(T)) = \\ 
		\int_0^\infty \prob(T_i^d\leq T+w|T_i^d>T, Z_i=z, H_i(T) = h(T), u_i=u)\cdot g(u|T^d_i>T, Z_i=z, H_i(T) = h(T))du \\
		= \frac{\int_0^\infty \left[ S_0^d(T)^{C_d u^\gamma} - S_0^d(T+w)^{C_d u^\gamma} \right]\cdot u^J \cdot S^r(T|h(T))^{C_r u}\cdot g(u)du}{\int_0^\infty S_0^d(T)^{C_d u^\gamma}\cdot u^J \cdot S^r(T|h(T))^{C_r u}\cdot g(u)du}.
	\end{aligned}
\end{equation}

%%%%%%%%%%%%%%%%%%%%%%%%%%%%%%%%%%%%%%%%%%%%%%%%%%%%%%%%%%%%%%%%%%%%%%%%%%
%%%%%%%%%%%%%%%%%%%%%%%%%%%%%%%%%%%%%%%%%%%%%%%%%%%%%%%%%%%%%%%%%%%%%%%%%%
\cleardoublepage	
\section{Distributional effect of hospitalizations in the renewal model}
\label{app:distribution_effect}

This appendix formally derives the results in Section~\ref{sec:renewal_dependence}. We start by studying how $S^r(T|h(T))$ varies with the distribution of hospitalizations. Fix the number of hospitalizations $J \geq 1$ and the follow-up time $T > 0$. Hospitalization times are distributed in the following set
\begin{equation}
	\mathcal{T} \equiv \{ (t_1, \dots, t_J) \in \mathbb{R}^J \colon 0 < t_1 < \cdots < t_J < T\}.
\end{equation}

The next result characterizes the behavior of $S^r(T|h(T))$ on this set. Recall that $\lambda_0^r$ is the baseline hazard for hospitalizations.
\begin{prop}{\label{prop:gap_distribution}}
	Suppose that $S^r$ is given by equation~\eqref{eq:surv_renewal}. If $\lambda_0^r$ is strictly increasing:
	\begin{itemize}
		\item $S^r(T|h(T))$ achieves a maximum value of
		\begin{equation}
			\left[S_0^r\left(\frac{T}{J+1}\right)\right]^{J+1}
		\end{equation}
		on $\mathcal{T}$. The maximum is achieved when $t_j=j\cdot T/(J+1)$ for $j\in{1,\dots, J}$.
		
		\item For every $(t_j)_{j=1}^{J} \in \mathcal{T}$, it holds that $S^r(T|h(T)) \geq S_0^r(T)$.
	\end{itemize}
	On the other hand, if $\lambda_0^r$ is strictly decreasing:
	\begin{itemize}
		\item $S^r(T|h(T))$ achieves a minimum value of
		\begin{equation}
			\left[S_0^r\left(\frac{T}{J+1}\right)\right]^{J+1}
		\end{equation}
		on $\mathcal{T}$. The minimum is achieved when $t_j=j\cdot T/(J+1)$ for $j\in{1,\dots, J}$.
		
		\item For every $(t_j)_{j=1}^{J} \in \mathcal{T}$, it holds that $S^r(T|h(T)) \leq S_0^r(T)$.
	\end{itemize}
\end{prop}
The proof of the above proposition is provided at the end of the appendix. To read the result, say that $\lambda_0^r$ is strictly increasing. Then the maximum value of $S^r(T|h(T))$ is achieved when hospitalizations are equiespaced on $[0, T]$. In this situation, hospitalizations are as dispersed as possible. On the contrary, one of the points where the minimum value of $S^r$ is achieved is $t_j=T$ for every $j=1,\dots, J$. This point is not feasible since hospitalizations cannot occur simultaneously. What Proposition~\ref{prop:gap_distribution} says is that, when $\lambda_0^r$ is strictly increasing, $S^r(T|h(T))$ tends to its minimum as hospitalizations become more concentrated (i.e., closer to the point $t_j=T$ for every $j$). Note that this conclusion holds as long as $S_0^r$ is continuous, which is generally the case.

To study how the risk of death depends on $S^r(T|h(T))$, we introduce an auxiliary function Define the function $\varphi\colon (0,1) \times (0,1) \times (0,1) \to \mathbb{R}$ as
\begin{equation}
	\varphi(x,y,v; J,\gamma) \equiv \frac{\int_0^\infty \left(x^{u^\gamma}-y^{u^\gamma}\right) \cdot v^{u}\cdot u^J \cdot g(u)du}{\int_0^\infty x^{u^\gamma} \cdot v^{u}\cdot u^J \cdot g(u)du}.
\end{equation}
The function depends on the parameters $\gamma$ and $J$. Recall that $\gamma$ measures the degree of dependence between the death and hospitalization process. Then, for a patient with $J$ hospitalizations and and covariates $Z_i=z$, the predicted risk of death is 
\begin{equation} \label{eq:prob_phi}
		\mathbb{P}(T, w |z, h(T))=\varphi\left(S_0^d(T)^{C_d}, S_0^d(T+w)^{C_d}, S^r(T| h(T))^{C_r}; J, \gamma\right).
\end{equation}

The next proposition characterizes the relationship between the risk of death, $\mathbb{P}(T, w |z, h(T))$, and the value of the survival function of hospitalizations given their, history $S^r(T|h(T))$.
\begin{prop}{\label{prop:deriv_z}}
	For $\gamma=0$, $\varphi$ is constant in $v$. Moreover, if $x>y$, $\partial\varphi/\partial v$ is positive for $\gamma>0$ and negative for $\gamma<0$.
\end{prop}
 We note that, if the baseline survival function for death is not constant in $[T, T+w]$, we have that $S_0^d(T) > S_0^d(T+w)$. That is, the above proposition can be applied to equation~\eqref{eq:prob_phi}.

The next theorem puts together Propositions \ref{prop:gap_distribution} and \ref{prop:deriv_z}.
\begin{thm} \label{thm:charac_renewal}
	In the renewal model, where $S^r$ is given by equation~\eqref{eq:surv_renewal}, we have that
	\begin{itemize}
		\item $\mathbb{P}(T, w |z, h(T))$ is highest on $\mathcal{T}$ for $t_j=j\cdot T/(J+1)$ (dispersed hospitalizations) when (i) $\gamma>0$ and $\lambda_0^r$ is increasing or (ii) $\gamma < 0$ and $\lambda_0^r$ is decreasing.
		
		\item $\mathbb{P}(T, w |z, h(T))$ is lowest on $\mathcal{T}$ for $t_j=j\cdot T/(J+1)$ (dispersed hospitalizations) when (i) $\gamma>0$ and $\lambda_0^r$ is decreasing or (ii) $\gamma < 0$ and $\lambda_0^r$ is increasing.
	\end{itemize}
\end{thm}

\subsection*{Proofs}

\begin{proofc}[Proposition~\ref{prop:gap_distribution}]
	Define the baseline cumulative hazard for hospitalizations as $\Lambda_0^r(t)\equiv \int_0^s \lambda_0^r(s)ds$, so that $S_0^r(t)=\exp\{-\Lambda_0^r(t)\}$.  Also, recall that $t_0 \equiv 0$ and $t_{J+1} \equiv T$ and let $w_j\equiv t_{j}-t_{j-1}$ for $j\in\{1,\dots, J+1\}$. We want to study the behavior of 
	\begin{equation}
		\mathcal{O}(w_1,\dots, w_{J+1})\equiv \prod_{j=1}^{J+1}S_0^r(w_j)
	\end{equation}
	as a function of gap times $(w_1,\dots, w_{J+1})$. The chosen of $w_j$'s must satisfy $\sum_{j=1}^{J+1}w_j=t_{J+1}-t_0=T$ and $w_j > 0$ for every $j$. Note that
	\begin{equation}
		-\log\mathcal{O}(w_1,\dots, w_{J+1}) = \sum_{j=1}^{J+1} \Lambda_0^r(w_j).
	\end{equation}
	
	We show the results for strictly increasing $\lambda_0^r$. The results for strictly decreasing baseline hazard follow \textit{mutatis mutandis}. If $\lambda_0^r$ is strictly increasing, then $\Lambda_0^r$ is a convex function. Therefore, by Jensen's inequality:
	\begin{equation}
		\frac{\sum_{j=1}^{J+1}\Lambda_0^r(w_j)}{J+1} \geq \Lambda_0^r\left(\frac{\sum_{j=1}^{J+1}w_j}{J+1}\right)=\Lambda_0^r\left(\frac{T}{J+1}\right)
	\end{equation}
	Moreover, equality is achieved if and only if $w_1=\cdots=w_{J+1}$. That is, when $w_j=T/(J+1)$ for every $j$. The inequality in the above display implies that
	\begin{equation}
		\begin{aligned}
				-\log\mathcal{O}(w_1,\dots, w_{J+1}) &\geq (J+1)\Lambda_0^r\left(\frac{T}{J+1}\right)\Leftrightarrow \\
				\mathcal{O}(w_1,\dots, w_{J+1}) &\leq \exp\left\{-(J+1)\Lambda_0^r\left(\frac{T}{J+1}\right)\right\}.
		\end{aligned}
	\end{equation}
The maximum is achieved when $w_j=T/(J+1)$ for every $j$. 

To show the condition $	\mathcal{O}(w_1,\dots, w_{J+1})\geq \exp\{-\Lambda_0^r(T)\}$, let us introduce the sets
\begin{equation}
	\mathcal{C} \equiv \left\{w\in\mathbb{R}^{J+1} \colon \sum_{j=1}^{J+1} \Lambda_0^r(w_j) \leq \Lambda_0^r(T)\right\} \text{ and } 	\mathcal{W} \equiv \left\{w\in\mathbb{R}^{J+1} \colon \sum_{j=1}^{J+1} w_j = T, w_j \geq 0 \right\}.	
\end{equation}
Note that the feasible $w_j$'s lie in set $\mathcal{W}$ ---precisely, $\mathcal{W}$ is the euclidean closure of the set of feasible $w_j$'s. We show that $\mathcal{W} \subseteq \mathcal{C}$. This concludes the proof since it implies that for every feasible $w_j$'s
\begin{equation}
	 \sum_{j=1}^{J+1} \Lambda_0^r(w_j) \leq \Lambda_0^r(T) \Leftrightarrow 	\mathcal{O}(w_1,\dots, w_{J+1}) \geq \exp\{-\Lambda_0^r(T)\}.
\end{equation}

First, since $\Lambda_0^r$ is convex, for $w,w'\in\mathcal{C}$ and $\eta\in[0,1]$:
\begin{equation}
	 \sum_{j=1}^{J+1} \Lambda_0^r(\eta w_j+(1-\eta)w_j') \leq \eta\sum_{j=1}^{J+1} \Lambda_0^r(w_j)+(1-\eta)\sum_{j=1}^{J+1} \Lambda_0^r(w_j') \leq \Lambda_0^r(T),
\end{equation}
so $\mathcal{C}$ is a convex set. Then, define $\tau^j\in\mathbb{R}^{J+1}$ as the vector which takes value $T$ at coordinate $j$ and 0 everywhere else: $\tau_i^j=T \iff i=j$ and $\tau_i^j=0 \iff i\neq j$. Note that $\tau^j\in\mathcal{C}$ for every $j$. Moreover, for every $w\in\mathcal{W}$, $w=\sum_{j=1}^{J+1}\eta_j\tau^j$ with $\eta_j\equiv w_j/T$. Since $\eta_j\geq 0$ for every $j$ and $\sum_{j=1}^{J+1}\eta_j=1$, this means that every $w\in\mathcal{W}$ may be written as a convex combination of $\tau^j$'s. Thus, $\tau^j$ being in $\mathcal{C}$ and $\mathcal{C}$ being convex implies $\mathcal{W}\subseteq \mathcal{C}$. 

\end{proofc}

The proof of Proposition~\ref{prop:deriv_z} depends on the following result:
	\begin{lma}\label{lma:expect_ineq}
		For a positive non-degenerate random variable $\xi$ it holds that
		\begin{itemize}
			\item $\E\left[\xi^{1+\gamma}\right] > \E\left[ \xi\right] \cdot \E\left[\xi^\gamma\right]$  if  $\gamma > 0 $ and
			\item  $\E\left[\xi^{1+\gamma}\right] < \E\left[ \xi\right] \cdot \E\left[\xi^\gamma\right]$  if $\gamma < 0 $.
		\end{itemize}
	\end{lma}
	\begin{proofd}
		Let $f(\gamma) \equiv \E\left[\xi^{1+\gamma}\right] - \E\left[ \xi\right] \cdot \E\left[\xi^\gamma\right]$. We have that
		\begin{equation}
			f(\gamma) = \E\left[\xi^\gamma \cdot \left(\xi - \E[\xi]\right)\right] = \E\left[\left(\xi^\gamma - \E\left[\xi\right]^\gamma \right)\cdot \left(\xi - \E[\xi]\right)\right]. 
		\end{equation}
		Consider the function $\phi(x)\equiv x^\gamma$ defined for $x>0$. If $\gamma>0$, $\phi$ is strictly increasing. Therefore
		\begin{equation}
			\xi > \E[\xi] \iff \xi^\gamma > \E[\xi]^\gamma \text{ almost surely}.
		\end{equation}
		Thus, $f$ is the expectation of a non-negative random variable. Since $\xi$ is non-degenerate, the expectation is strictly positive: $f(\gamma)>0$.
		
		If $\gamma<0$, $\phi$ is strictly decreasing and thus $\xi > \E[\xi] \iff \xi^\gamma < \E[\xi]^\gamma$ almost surely. Then, $f$ is the expectation of a non-positive random variable.  Since $\xi$ is non-degenerate, the expectation is strictly negative: $f(\gamma)<0$.
		
	\end{proofd}

	\begin{proofc}[Proposition~\ref{prop:deriv_z}]
		First, when $\gamma=0$, $\varphi$ does not depend on $v$:
		\begin{equation}
			\varphi(x,y,v; J,0) \equiv \frac{\int_0^\infty \left(x - y\right) \cdot v^{u}\cdot u^J \cdot g(u)du}{\int_0^\infty x \cdot v^{u}\cdot u^J \cdot g(u)du} = \frac{x-y}{x}.
		\end{equation}
		
		For $\gamma\neq0$, taking derivatives of $\varphi$ with respect to $v$:
		\small{
			\begin{equation}
				\begin{aligned}
					\frac{\partial\varphi}{\partial v}(x,y,v) &= \frac{1}{v \cdot \left(\int_0^\infty u^J \cdot x^{u^\gamma} \cdot v^{u}\cdot  g(u)du\right)^2} \left[\int_0^\infty u^{J+1} \cdot x^{u^\gamma} \cdot v^{u}\cdot g(u)du \cdot \int_0^\infty u^{J} \cdot y^{u^\gamma} \cdot v^{u}\cdot g(u)du \right. \\
					& \left. - \int_0^\infty u^{J+1} \cdot y^{u^\gamma} \cdot v^{u}\cdot g(u)du \cdot \int_0^\infty u^{J} \cdot x^{u^\gamma} \cdot v^{u}\cdot g(u)du \right].
				\end{aligned}
		\end{equation}}
		Define
		\begin{equation}
			F(x)\equiv \frac{\int_0^\infty u^{J+1} \cdot x^{u^\gamma} \cdot v^{u}\cdot g(u)du}{\int_0^\infty u^{J} \cdot x^{u^\gamma} \cdot v^{u}\cdot g(u)du}.
		\end{equation}
		Rearranging terms in the numerator of $\partial\varphi/\partial v$ leads to
		\begin{equation}
				\frac{\partial\varphi}{\partial v}(x,y,v) > 0 \iff F(x) > F(y) \text{ and } \frac{\partial\varphi}{\partial v}(x,y,v) < 0 \iff F(x) < F(y).
		\end{equation}
		Since $x>y$, it suffices to check whether $F$ is strictly increasing or decreasing, depending on the sign of $\gamma$.  Taking derivatives:
		\begin{equation}
			\begin{aligned}
				\frac{dF}{dx}(x) &= \frac{1}{x \cdot \left(\int_0^\infty u^J \cdot x^{u^\gamma} \cdot v^{u}\cdot g(u)du\right)^2} \left[\int_0^\infty u^{J+1+\gamma } \cdot x^{u^\gamma} \cdot v^{u}\cdot g(u)du \cdot \int_0^\infty u^{J} \cdot x^{u^\gamma} \cdot v^{u}\cdot g(u)du \right. \\
				& \left. - \int_0^\infty u^{J+1} \cdot x^{u^\gamma} \cdot v^{u}\cdot g(u)du \cdot \int_0^\infty u^{J+\gamma} \cdot x^{u^\gamma} \cdot v^{u}\cdot g(u)du \right].
			\end{aligned}
		\end{equation}
		
		Let $m_\gamma(u)\equiv u^Jx^{u^\gamma}v^ug(u)$. The sign of the derivative of $F$ depends on the sign of the numerator
		\begin{equation}
			\operatorname{NUM}(\gamma)\equiv \int_0^\infty u^{1+\gamma}m_\gamma(u)du \cdot \int_0^\infty m_\gamma(u)du - \int_0^\infty u m_\gamma(u)du \cdot \int_0^\infty u^\gamma m_\gamma(u)du
		\end{equation}	
		Let $M_\gamma \equiv  \int_0^\infty m_\gamma(u)du > 0$ and define the pdf $p_\gamma(u) \equiv m_\gamma(u)/M_\gamma$. We then have that
		\begin{equation}
			x \cdot \frac{dF}{dx}(x) = \frac{\operatorname{NUM}(\gamma)}{M_\gamma^2} = \int_{0}^{\infty} u^{1+\gamma}p_\gamma(u)du - \int_{0}^{\infty} u p_\gamma(u)du \cdot \int_{0}^{\infty} u^{\gamma}p_\gamma(u)du.
		\end{equation}
		We can then apply Lemma~\ref{lma:expect_ineq} to a random variable $\xi$ with density $p_\gamma$ to get the result (recall that $x>0$).
	\end{proofc}

	\begin{proofc}[Theorem~\ref{thm:charac_renewal}]
		Suppose that $\gamma > 0$. By Proposition~\ref{prop:deriv_z} and $C_r > 0$,
		\begin{equation}
			v \mapsto \varphi\left(S_0^d(T)^{C_d}, S_0^d(T+w)^{C_d}, v^{C_r}; J, \gamma\right)
		\end{equation}
		is a strictly increasing transformation. Thus, the maximizer and minimizer of $\mathbb{P}(T, w |z, h(T))$ on $\mathcal{T}$ coincide with the maximizer and minimizer of $S^r(T|h(T))$ on $\mathcal{T}$. By Proposition~\ref{prop:gap_distribution}, $t_j=j\cdot T/(J+1)$ is a maximizer when $\lambda_0^r$ is increasing and a minimizer when $\lambda_0^r$ is decreasing.

		If $\gamma < 0$, Proposition~\ref{prop:deriv_z} states that
		\begin{equation}
			v \mapsto \varphi\left(S_0^d(T)^{C_d}, S_0^d(T+w)^{C_d}, v^{C_r}; J, \gamma\right)
		\end{equation}
		is a strictly decreasing transformation. This means that a maximizer of $S^r(T|h(T))$ on $\mathcal{T}$ is a minimizer of $\mathbb{P}(T, w |z, h(T))$ on $\mathcal{T}$. Also,  a minimizer of $S^r(T|h(T))$ on $\mathcal{T}$ is a maximizer of $\mathbb{P}(T, w |z, h(T))$ on $\mathcal{T}$. The conclusion for this case follows by applying Proposition~\ref{prop:gap_distribution} again.
	\end{proofc}

\end{appendices}

\newpage
\addcontentsline{toc}{section}{References}
\makeatletter
\makeatother
\bibliographystyle{unsrtnat-AMA} % sorted by order of appearance and AMA style
\bibliography{references}

\end{document}